\documentclass[traditabstract]{aa} 
\usepackage{graphicx}
\usepackage{natbib}
\usepackage{longtable,lscape}
\usepackage{txfonts}

\usepackage{float}

\def\mstar  {$M_{\star}$}
\def\macc   {$\dot{M}_{\rm acc}$}
\def\lacc   {$L_{\rm acc}$}

\def\msun {$M_{\odot}$}
\def\lsun {$L_{\odot}$}
\def\lstar {$L_\star$}

\def\lline {$L_{\rm line}$}

\newcommand{\Ll}{{$L_{\rm line}$}}

\def\iso {ISO$-$Oph}

\defcitealias{Natta06}{N06}
\defcitealias{Natta04}{N04}

\begin{document}

   \title{X-Shooter study of accretion in $\rho$-Ophiucus: \\very low-mass stars and brown dwarfs \thanks{This work is based on observations made with ESO Telescopes at the Paranal Observatory under programme ID  085.C-0876. }}

   \author{C.F. Manara\inst{1}\fnmsep\thanks{ESA Research Fellow}, 
        L. Testi\inst{2,3,4}, A. Natta\inst{3,5}, \and J. M. Alcal\'{a}\inst{6}
          }
          
     \authorrunning{C.F. Manara et al.}
     \titlerunning{X-Shooter study of accretion in $\rho$-Ophiucus}

 \institute{Scientific Support Office, Directorate of Science and Robotic Exploration, European Space Research and Technology Centre (ESA/ESTEC), Keplerlaan 1, 2201 AZ Noordwijk, The Netherlands\\
               \email{cmanara@cosmos.esa.int}
                \and
         European Southern Observatory, Karl Schwarzschild Str. 2, 85748 Garching bei M{\"u}nchen, Germany
         \and
             INAF/Osservatorio Astrofisico of Arcetri, Largo E. Fermi, 5, 50125 Firenze, Italy
        \and
                Excellence Cluster Universe, Boltzmannstr. 2, D-85748 Garching bei M{\"u}nchen, Germany
         \and
             School of Cosmic Physics, Dublin Institute for Advanced Studies, 31 Fitzwilliams Place, 2 Dublin, Ireland
         \and
             INAF/Osservatorio Astronomico di Capodimonte, Salita Moiariello, 16 80131 Napoli, Italy
}

   \date{Received March, 24th 2015; accepted May, 4th 2015}

 
  \abstract{We present new VLT/X-Shooter optical and near-infrared spectra of a sample of 17 candidate young low-mass stars and brown dwarfs located in the $\rho$-Ophiucus cluster. We  derived the spectral type and extinction for all the targets, and then we determined their physical parameters. All the objects but one have \mstar$\lesssim$0.6 \msun, and eight have mass below or close to the hydrogen-burning limit. Using the intensity of various permitted emission lines present in their spectra, we determined the accretion luminosity and mass accretion rates (\macc) for all the objects. When compared with previous works targeting the same sample, we find that, in general, these objects are not as strongly accreting as previously reported, and we suggest that the reason is our more accurate estimate of the photospheric parameters. We also compare our findings with recent works in other slightly older star-forming regions, such as Lupus, to investigate possible differences in the accretion properties, but we find that the accretion properties for our targets have the same dependence on the stellar and substellar parameters as in the other regions. This leads us to conclude that we do not find evidence for a different dependence of \macc \ with \mstar \ when comparing low-mass stars and brown dwarfs. Moreover, we find a similar small ($\lesssim$1 dex) scatter in the \macc-\mstar \ relation as in some of our recent works in other star-forming regions, and no significant differences in \macc \ due to different ages or properties of the regions. The latter result suffers, however, from low statistics and sample selection biases in the current studies. The small scatter in the \macc-\mstar \ correlation confirms that mass accretion rate measurements in the literature based on uncertain photospheric parameters and single accretion indicators, such as the H$\alpha$ width, can lead to a scatter that is unphysically large. 
Our studies show that only broadband spectroscopic surveys coupled with a detailed analysis of the photospheric and accretion properties  allows us to properly study the evolution of disk accretion rates in star-forming regions. 

}


 \authorrunning{Manara et al.}
\titlerunning{X-Shooter study of accretion in $\rho$-Ophiucus}
\maketitle

%

\section{Introduction}\label{sect::intro}

The evolution of protoplanetary disks surrounding forming stars has been a major subject of study in recent years. The interest in this topic is driven by the fact that disks are the birthplace of planets. A clear understanding of the physical mechanisms driving the formation, evolution, and dispersal of disks is thus needed to constrain planet formation theories, and to explain the observed properties of our own solar system and of the wealth of exoplanetary systems discovered so far.

During their evolution, the central young stellar object (YSO) and the surrounding disk interact through multiple processes, such as photoevaporation, stellar winds, and accretion of matter. The latter is a result of viscous processes happening in the disk, which drive its secular evolution \citep[e.g.,][]{Hartmann98}, while photoevaporation and winds/outflows are thought to play a major role in the final disk dispersal \citep[e.g.,][and references therein]{Alexander13}. A good understanding of these processes is needed to describe how disks evolve, and thus how planet formation takes place. Star-disk interaction processes can be studied observationally through the strong signatures they introduce in the spectra of YSOs. Accretion shocks give rise to continuum excess emission in the UV \citep[e.g.,][]{Valenti93,Gullbring98,Gullbring00,Calvet98,Calvet00} and the prominent emission of permitted lines across the whole spectrum \citep[e.g.,][]{Muzerolle98,Muzerolle98b,Muzerolle98c,Muzerolle03,Natta04}, while winds are traced by various forbidden emission lines \citep[e.g.,][]{Hartigan95,Rigliaco13,Natta14}. In  recent years, new instruments have provided the possibility of studying  these processes simultaneously in large samples of objects. In particular, the X-Shooter spectrograph mounted on the ESO/VLT telescope \citep{Vernet11} provides us the possibility of studying YSOs with unprecedented levels of detail, as it simultaneously takes  spectra at medium resolution (R$\sim$5000-20000) over the complete wavelength range from $\sim$300 nm to $\sim$2500 nm. We are thus able to study  all the signatures of accretion and winds in the optical and near-infrared spectra of YSOs at the same time. 

A significant effort has been made recently to study YSOs with X-Shooter. The accretion process has been studied for objects located in various nearby star-forming regions ($\sigma$-Ori, \citealt{Rigliaco12}; Lupus, \citealt{Alcala14}; Chamaeleon, \citealt{ManaraChaI}) and at various phases of disk evolution \citep[e.g., transitional disks;][]{Manara14}. These works, among various results, showed that the mass accretion rates (\macc) dependence on the stellar mass (\mstar) has a significant smaller scatter ($\sim$0.4 dex) than previously observed \citep{Alcala14}, and that there are transitional disks with comparable accretion rates as less evolved disks \citep{Manara14}. 
These works focused on very low-mass, pre-main-sequence (PMS) stars and up to solar mass YSOs, with only a few objects below the hydrogen burning limit \citep[e.g.,][]{Rigliaco11b, Stelzer13}. These objects were located in regions with age $\sim$2-3 Myr, or more. However, to fully test models of disk evolution it is important to derive accretion properties of YSOs for a large range of stellar masses and age. The work we present here aims to enlarge the sample of very low-mass stars (VLMS) and brown dwarfs (BDs) studied with X-Shooter, focusing on objects located in the younger (age$\sim$1 Myr) $\rho$-Oph embedded cluster.

The $\rho$-Oph cluster is located at a distance $d\sim$ 125 pc \citep{Lombardi08,Loinard08}, and it is still highly embedded in the parental cloud, with values of $A_V$ up to 50 mag in the densest core \citep{Wilking83}, and with a rather high value of total-to-selective extinction ratio $R_V\sim$ 5.6 \citep[e.g.,][]{Kenyon98,Chapman09,McClure10,Comeron10}. The total population of the surrounding Ophiucus region is estimated to be of around 300 YSOs \citep{Evans09}, with $\sim$200 YSOs located in $\rho$-Oph itself. Given the large quantity of YSOs present in this cluster and its proximity, it has been extensively studied in the past at various wavelengths from optical, near- and far-infrared, millimeter, radio, to X-ray \citep[e.g.,][]{Greene95,Luhman99,Wilking05,AlvesdeOliveira10, AlvesdeOliveira12, AlvesdeOliveira13, Bontemps01,McClure10,Gagne04}. 

Various works aimed to understand the accretion properties of YSOs at early ages have targeted $\rho$-Oph. In particular, \citet[hereafter N06]{Natta06} have studied the vast majority of the Class~II YSOs in this cluster, selected from the ISO sample of \citet{Bontemps01}. They used near-infrared emission lines (Pa$\beta$ and Br$\gamma$) to determine the accretion rates, and a statistical approach, based on the assumption that the whole population is located on the Hertzprung-Russel diagram (HRD) on a single isochrone at 0.5 Myr, to derive stellar parameters for most of the targets\footnote{Their results were obtained using an old distance estimate of 160 pc, and have been corrected for the newly determined distance and recalculated with the same method, but assuming a single isochrone at 1 Myr for the whole population by \citet{Rigliaco11a}. Here we report in Tables~\ref{tab::natta_125} all the values derived using the most recent distance estimated for the cluster of 125 pc.}. This work showed that accretion rates for BDs in $\rho$-Oph are typically higher than for objects with the same mass in Taurus, and that various strongly accreting BDs are found in this region. However, this result strongly depends on the already mentioned assumptions, which can result in incorrect estimates of the stellar parameters, extinction, or accretion for single objects, and only more precise analyses can shed light on this issue. With the set of X-Shooter spectra we present here, we are able to derive stellar and accretion parameters with significantly higher precision. With this in hand, we aim at addressing the still open questions comparing accretion properties for BDs and VLMS in this region and in other samples observed with X-Shooter.

The paper is organized as follows. In Section~\ref{sect::obs} we describe the sample, observations, and  data reduction; in Section~\ref{sect::method} we then discuss  the method used to analyze the spectra and the derived properties of the objects, together with their main spectral features. In Section~\ref{sect::acc_prop} we determine the accretion properties of the targets, and discuss the differences with previous estimates for the same targets;  the comparison of accretion properties with results from other regions is then carried out in Section~\ref{sect::acc_comp}. 
Finally, we discuss our findings and conclusions in Section~\ref{sect::conclusions}.

\begin{table*}
\begin{center}
\footnotesize
\caption{\label{reg::tab::oph_targets} Sample, observing log, and SNR of the spectra}
\begin{tabular}{l|cc|c|c|ccc}
\hline \hline
 Object/other name &  RA(2000)  & DEC(2000) &  Obs. date  & Exp. Time & \multicolumn{3}{c}{SNR at $\lambda$=} \\ 
      &  h \, :m \, :s & $^\circ$ \, ' \, ''   &  YY-MM-DD & [s]  & 700 & 855& 1300 \\         
\hline

ISO$-$Oph023  /  SKS1   &  16:26:18.821  &  $-$24:26:10.52    & 2010-08-20 & 4$\times$750s                         &   0  &  4  & 61                       \\
ISO$-$Oph030  /  GY5   &  16:26:21.528  &  $-$24:26:00.96    & 2010-08-20 & 4$\times$750s                                 &  15  & 71  & 66               \\
ISO$-$Oph032  /  GY3   &  16:26:21.899  &  $-$24:44:39.76    & 2010-06-01 & 4$\times$750s                                 &  12  & 29  & 65               \\
ISO$-$Oph033  /  GY11   &  16:26:22.269  &  $-$24:24:07.06    & 2010-08-26 & 4$\times$1800s                        &   0  &  2  & 29                       \\
ISO$-$Oph037  /  LFAM3  /  GY21   &  16:26:23.580  &  $-$24:24:39.50    & 2010-08-21 & 4$\times$750s      &   0  &  2  & 36                                       \\
ISO$-$Oph072  /  WL18    &  16:26:48.980  &  $-$24:38:25.24    & 2010-04-06 & 4$\times$480s                         &   3  & 17  & 51                       \\
ISO$-$Oph087     &  16:26:58.639  &  $-$24:18:34.66    & 2010-08-28 & 4$\times$750s                                         &  0  &  2  & 29        \\
ISO$-$Oph094     &  16:27:03.591  &  $-$24:20:05.45    & 2010-08-28 & 4$\times$750s                                         &   0  &  1  & 10       \\
ISO$-$Oph102  /  GY204    &  16:27:06.596  &  $-$24:41:48.84    & 2010-07-23 & 4$\times$750s                         &  12  & 41  & 68                       \\
ISO$-$Oph115  /  WL11  /  GY229   &  16:27:12.131  &  $-$24:34:49.14    & 2010-07-30 & 4$\times$750s      &   0  &  1  & 37                                       \\
ISO$-$Oph117  /  WLY2-32b  /  GY235   &  16:27:13.823  &  $-$24:43:31.66    & 2010-06-08 & 4$\times$480s &   3 &  17 &  92                                               \\
ISO$-$Oph123     &  16:27:17.590  &  $-$24:05:13.70    & 2010-08-27 & 4$\times$750s                                         &  58  & 74  & 83       \\
ISO$-$Oph160  /  B162737-241756    &  16:27:37.422  &  $-$24:17:54.87    & 2010-08-27 & 4$\times$750s    &   1  & 15  & 62                                       \\
ISO$-$Oph164  /  GY310   &  16:27:38.631  &  $-$24:38:39.19    & 2010-06-08 & 4$\times$750s                         &   6  & 31  & 45                       \\
ISO$-$Oph165  /  GY312   &  16:27:38.945  &  $-$24:40:20.67    & 2010-06-08 & 4$\times$750s                         &   0  &  1  & 27                       \\
ISO$-$Oph176  /  GY350   &  16:27:46.291  &  $-$24:31:41.19    & 2010-08-21 & 4$\times$750s                         &   1  & 16  & 89                       \\
ISO$-$Oph193  /  B162812-241138   &  16:28:12.720  &  $-$24:11:35.60    & 2010-08-21 & 4$\times$750s      &   2  & 28  & 24                                       \\

\hline

\end{tabular}
\tablefoot{ Exposure times are the same in the three X-Shooter arms (UVB, VIS, and NIR). Wavelengths for the SNR calculation are reported in nm.}

\end{center}
\end{table*}


\section{Sample, observations, and data reduction}\label{sect::obs}

We report here about new data collected with the ESO/VLT X-Shooter spectrograph \citep{Vernet11} during Pr.Id.085.C-0876 (PI Testi). This instrument acquires spectra of the targets from $\lambda\sim$300 nm to $\lambda\sim$2500 nm simultaneously, splitting the spectrum in three parts, usually referred to as the UVB ($\lambda\lambda\sim$300-550 nm), VIS ($\lambda\lambda\sim$550-1050 nm), and NIR ($\lambda\lambda\sim$1050-2500 nm) arms.

The sample was selected from the one of \citetalias{Natta06}  to cover as many of the objects with detected accretion and estimated stellar mass \mstar$<$0.1 \msun \ as possible. It comprises 16 Class~II YSOs and one Class~I YSO that are analyzed here for the first time, one transitional disk (\iso196) that has been analyzed in \citet{Manara14}, and three Class~III YSOs. We focus on the analysis of the 16 Class~II YSOs and the Class~I YSO. All these targets were included in the ISO sample compiled by \citet{Bontemps01} and most of them have previously been  studied spectroscopically \citep[e.g.,][see also Table~\ref{tab::lit} for literature data on these targets]{Natta02,Natta04,Natta06,Wilking99,Wilking05}. Their ISO number, other names, and coordinates are reported in the first three columns of Table~\ref{reg::tab::oph_targets}. 

All the objects have been observed in service mode using the 1.0x11\arcsec \ slit in the UVB arm and the 0.9x11\arcsec \ slits in the VIS and NIR arms, which lead to a nominal resolution R = $\lambda$/$\Delta\lambda$$\sim$ 4350, 7450, and 5300 in the three arms, respectively. Different exposure times have been adopted for each target depending on their estimated fluxes. The dates of the observations and  exposure times are reported in the fourth and fifth columns of Table~\ref{reg::tab::oph_targets}. The late spectral type (SpT) of most of the targets and the high extinction of the region explain the low SNR of most spectra, at least in the UVB arm. The only spectra with any detected signal in the continuum in the UVB arm are those of ISO-Oph032 and ISO-Oph123, although with low SNR. Some objects have also very low SNR in the VIS spectra at $\lambda\lesssim$ 600-700 nm, or even at longer wavelengths. The NIR spectra of all targets have a good SNR. We report the SNR for the various spectra at different wavelengths in the last three columns of Table~\ref{reg::tab::oph_targets}.The object \iso072 was reported by \citet{McClure10} to be member of a binary system with a separation of 3.62\arcsec. The X-Shooter spectrum analyzed here is the one of the primary component of the system, while the secondary was not included in the slit.

Data reduction has been carried out using the X-Shooter pipeline \citep{Modigliani} version~1.3.7 and the same procedure as in \citet{Alcala14}. 
For the targets observed on 2010-06-01, 2010-08-20, and 2010-08-28, the photometric standard observed during the night of observation is not supported anymore in the pipeline. The flux calibration of the spectra has thus been obtained using the photometric standard star observed  the following night, which was always photometric.
The flux calibration of the spectra reduced with the pipeline has been compared with available 2MASS photometry and rescaled to this using synthetic photometry on the spectra to match the photometric flux. The correction factors are larger than 1 and usually lower than $\sim$1.5, and always less than $\sim$1.9. Indeed, these spectra have all been obtained with slits whose sizes are comparable with the seeing, so the slit losses are small. The only spectrum in which the conjunctions between spectra from different arms are not very good is ISO-Oph102. In this object, the quality of the spectrum in the last $\sim$100 nm of the VIS arm is not good, and this results in a bad matching of the VIS and NIR arms. Finally, telluric line removal has been performed using the IRAF\footnote{IRAF is distributed by National Optical Astronomy Observatories, which is operated by the Association of Universities for Research in Astronomy, Inc., under cooperative agreement with the National Science Foundation.} task \textit{telluric} following the procedure discussed in \citet{Alcala14} and using standard telluric spectra obtained close in time and air mass to the targets.


\section{Stellar and substellar properties}\label{sect::method}

Here we discuss the method adopted to derive the stellar and substellar properties of the targets, and also the results obtained and main features present in the spectra.

\subsection{Method}
Spectral classification for objects located in $\rho$-Oph is a difficult process, as these are usually highly-extincted and large excess because of disk emission, is also present  at near-IR wavelengths \citep[e.g.,][]{Luhman99,Wilking05,Natta06}. Thanks to the very large wavelength coverage of our spectra, however, we can derive the spectral type (SpT) and extinction ($A_V$) of the targets using multiple spectral features simultaneously, thus reducing the degeneracies present in this analysis. Given that the spectra have very low or zero SNR in the UVB arm, the analysis cannot be performed using the method described in \citet{Manara13b}, which simultaneously determines  SpT, $A_V$, and accretion luminosity (\lacc), fitting the UV and visible part of the spectra with a grid of models that includes the contribution of photospheric and accretion-induced emission, and reddening. Similarly, the analysis methods used by \citet{Alcala14} or \citet{Rigliaco12} cannot be adopted here, as all these need to perform a fit of the excess in the UV part of the spectrum to derive \lacc, and are well suited for the SpT determination only for objects with low extinction.

\begin{table}
\begin{center}
\footnotesize
\caption{\label{tab::grid} Grid of photospheric templates}
\begin{tabular}{l|cc|l}
\hline \hline
 Object &  SpT & T$_{\rm eff}$ [K] & Ref \\         
\hline

TWA9A & K5 & 4350       & 1 \\
TWA6            & K7 & 4060     & 1 \\
TWA25 & M0 & 3850 & 1 \\
TWA14   &       M0.5 & 3780 & 1 \\
TWA13B & M1 &3705 & 1 \\
synthetic spectrum & M1.5  & 3600 & 2 \\
Sz122   &       M2      & 3560 & 1 \\
synthetic spectrum & M2.5 &3500 & 2 \\
TWA7    &       M3      & 3415 &1\\
TWA15A  &       M3.5  & 3340 & 1 \\
Sz94    & M4    & 3270 & 1 \\
SO797   &       M4.5 &3200 & 1 \\
Par$-$Lup3$-$2  & M5 &3125      & 1 \\
SO999   &       M5.5 &  3060 & 1 \\
synthetic spectrum & M6  & 3000 & 2 \\
Par$-$Lup3$-$1 & M6.5  & 2935 & 1 \\
synthetic spectrum & M7 & 2900 & 2 \\
synthetic spectrum & M7.5  & 2800 & 2 \\
synthetic spectrum & M8  & 2700 & 2 \\
synthetic spectrum & M8.5  & 2550 & 2 \\
TWA26   &       M9      & 2400 & 1 \\
TWA29   & M9.5 & 2330   & 1 \\
synthetic spectrum & L0  & 2200 & 2 \\

\hline

\end{tabular}
\tablebib{1.~\citet{Manara13a}; 2.~\citet{Allard11}}

\end{center}
\end{table}

The procedure we use here is the following: We collect a grid of photospheric templates, which cover the SpT from K5 to L0, with a typical step of 0.5 spectral subclasses throughout the entire M subclass. These are chosen primarily from the Class~III spectra of \citet{Manara13a}, which are observed spectra of nonaccreting PMS stars obtained with the same instrument. As this grid of templates is incomplete for SpT later than M6.5 and earlier than M9, and has no spectra with SpTs M1.5, M2.5, and M6, to complete our grid we make use of the synthetic spectra BT-Settl of \citet{Allard11}, which are smoothed to match the resolution of the X-Shooter spectra. We adopt synthetic spectra with values of log$g$=3.5 (in cgs units), typical of young objects in $\rho-$Oph \citep[e.g.,][]{Comeron10}, and effective temperatures ($T_{\rm eff}$), corresponding to the given SpT according to the SpT-$T_{\rm eff}$ relation of \citet{Luhman03}. The details of the adopted grid of templates is reported in Table~\ref{tab::grid}. 
The synthetic spectra reproduce well, in general, the observed templates down to $T_{\rm eff}\sim$3050 K, i.e., SpT$\sim$M5.5, while there are some discrepancies at lower temperatures. This, unfortunately, is also the region of our observed template library, which has a sparse sampling of spectral types. We describe the typical reddening toward the objects with the reddening law of \citet{Cardelli} using a value of $R_V$=5.6. 

For each target, we find the stellar parameters (SpT, $A_V$) with an automatic procedure that compares the observed spectrum to the various templates, which are artificially reddened ($A_V$ ranges from 0 to 18 mag in step of 0.1 mag) and then normalized to the observed spectrum at $\lambda\sim$1025 nm. The comparison is done between the value of the mean flux across a window of $\sim$4 nm in 28 different points of the spectrum from $\lambda\sim$ 700 nm to $\lambda\sim$1725 nm, which cover various molecular features particularly strong in the spectra of VLMS and BDs and sensitive to $T_{\rm eff}$, such as TiO, VO, and other molecular features \citep[see][for a list]{Manara13a} in the VIS arm, and H$_2$O bands in the NIR arm. We limit our comparison to points within the aforementioned spectral range to avoid regions with large veiling due to accretion and low SNR spectra due to high extinction ($\lambda<$700 nm), and regions where the contribution of the disk emission is substantial ($\lambda\gtrsim$1700 nm). The best fit is derived by minimizing a $\chi^2_{\rm like}$ distribution, similar to the procedure by \citet{Manara13b}, on the two free parameters SpT and $A_V$. This is defined as

\begin{equation}
\chi^2_{\rm like} = \sum_{i} \left( \frac{f_{obs}(\lambda_i) - f_{mod}(\lambda_i)}{\sigma_{obs}(\lambda_i)} \right)^2,
\end{equation}

where the index $i$ denotes the various points included in the analysis, $f_{obs}$ is the flux of the observed spectrum, $f_{mod}$ that of the template, and $\sigma_{obs}$ is the standard deviation on the flux of the observed spectrum. Uncertainties in the estimated parameters (SpT, $A_V$) are derived from the $\Delta\chi^2_{\rm like}$ distribution, considered as a $\Delta\chi^2$ distribution. Points where the SNR on the observed target is less than 5 are excluded from the $\chi^2_{\rm like}$ calculation. To have enough SNR in the VIS arm, we use spectra smoothed using the \textit{boxcar smoothing} procedure included in the IRAF \textit{splot} package. This procedure convolves the spectrum with a rectangular box, whose width we set  to 7 pixels, and results in a broadening of the narrower absorption and emission lines and features, but preserves their fluxes and equivalent widths. The results of the fit do not change using the smoothed spectra, but only the uncertainties on the results and the numbers of points included in the fit. At the same time, the different resolution of the templates does not influence the result of the fit, as all the points where the $\chi^2_{\rm like}$ is calculated are free of narrow absorption lines. We also include a constraint on the accepted fits, namely that the observed spectrum should have a larger flux than the template at $\lambda>$2115 nm, i.e., in the K-band, a part of the spectrum not used otherwise in the analysis.

Various checks have been carried out on this procedure to determine its reliability, which we list here: $i$) Starting from our own grid of templates, we artificially add to their spectra various amounts of veiling due to accretion, of emission from the disk, and of reddening due to extinction. By applying the procedure just described in most cases we retrieve their correct stellar parameters. These stellar parameters only have  big discrepancies  in cases where the veiling due to accretion is substantial, i.e., the ratio of accretion to stellar luminosity (\lacc/\lstar) is larger than 0.5. This is expected to be the case for very few low-mass Class~II YSOs \citep[e.g.,][]{Alcala14}. $ii$) We fit several X-Shooter spectra of Class~II YSOs in Lupus by \citet{Alcala14} and retrieve results compatible with theirs, similarly to case $i$ , as these objects are not strongly veiled. $iii$) We run our procedure using only synthetic spectra for the later type objects or varying the resolution of the synthetic spectra. These different grids generally lead  to differences in the derived parameters well within their uncertainties. $iv$) We try to use spectra smoothed both in the VIS and NIR arms, while in general we smoothed only the VIS arm, and we obtain small differences. $v$) We try to use only non-smoothed spectra, with which we get similar results, but with larger uncertainties.

The method just described allows us to derive SpT and $A_V$ for all targets and lead to good best fits. We show in Appendix~\ref{app::plot_spectra} (Figs.~\ref{reg::fig::best_fit_oph}~$-$~\ref{reg::fig::best_fit_oph6}) the reddening corrected spectra with their best-fit template, also including  additional \textit{Spitzer} photometry from the literature. The values of $\chi^2_{\rm like, red} = \chi^2_{\rm like}/d.o.f.$ range from $\sim$0.5 to $\sim$3, with only two objects, ISO$-$Oph102 and ISO$-$Oph123, with $\chi^2_{\rm like, red}>$4 (see Appendix~\ref{app::ind_targ} for further discussion on these objects). We adopt a slightly different approach for the two objects with larger veiling due to accretion in our sample, ISO$-$Oph072 and ISO$-$Oph123. For these two targets we exclude from the $\chi^2_{\rm like}$ calculation all the points at $\lambda>$1340 nm, i.e., in the $H-$band. This choice is made as we expect that strongly accreting objects make a significant contribution to the emission at near-IR wavelengths from a disk, which is not included in our modeling. With this choice, we obtain the same result for ISO$-$Oph123 as with the inclusion of all the points in the $\chi^2_{\rm like}$ calculation, but with a lower value of $\chi^2_{\rm like, red}$. As we discuss in the following, the results for \iso123 are highly uncertain and the derived stellar properties for this object should be used with caution. For  ISO$-$Oph072, instead we derive a significantly later SpT (M3.5 instead of K7), but with a significantly smaller value of $\chi^2_{\rm like, red}$ (2.2 instead of 4.7). We check the results also  excluding the $H$-band  for objects with strong infrared excess due to an envelope, such as ISO-Oph037 and ISO-Oph165 (see Table~\ref{tab::lit} and Figs.~\ref{reg::fig::best_fit_oph}-\ref{reg::fig::best_fit_oph6}). For both of these targets, we derive $A_V$ and SpT within the uncertainties of the normal procedure including all the points. However, as fewer points are included in the fit, the statistical significance of the results is lower. We decide then to use the values obtained using all points in the fit for these two objects, as well.

\begin{figure}[]
\centering
\includegraphics[width=0.45\textwidth]{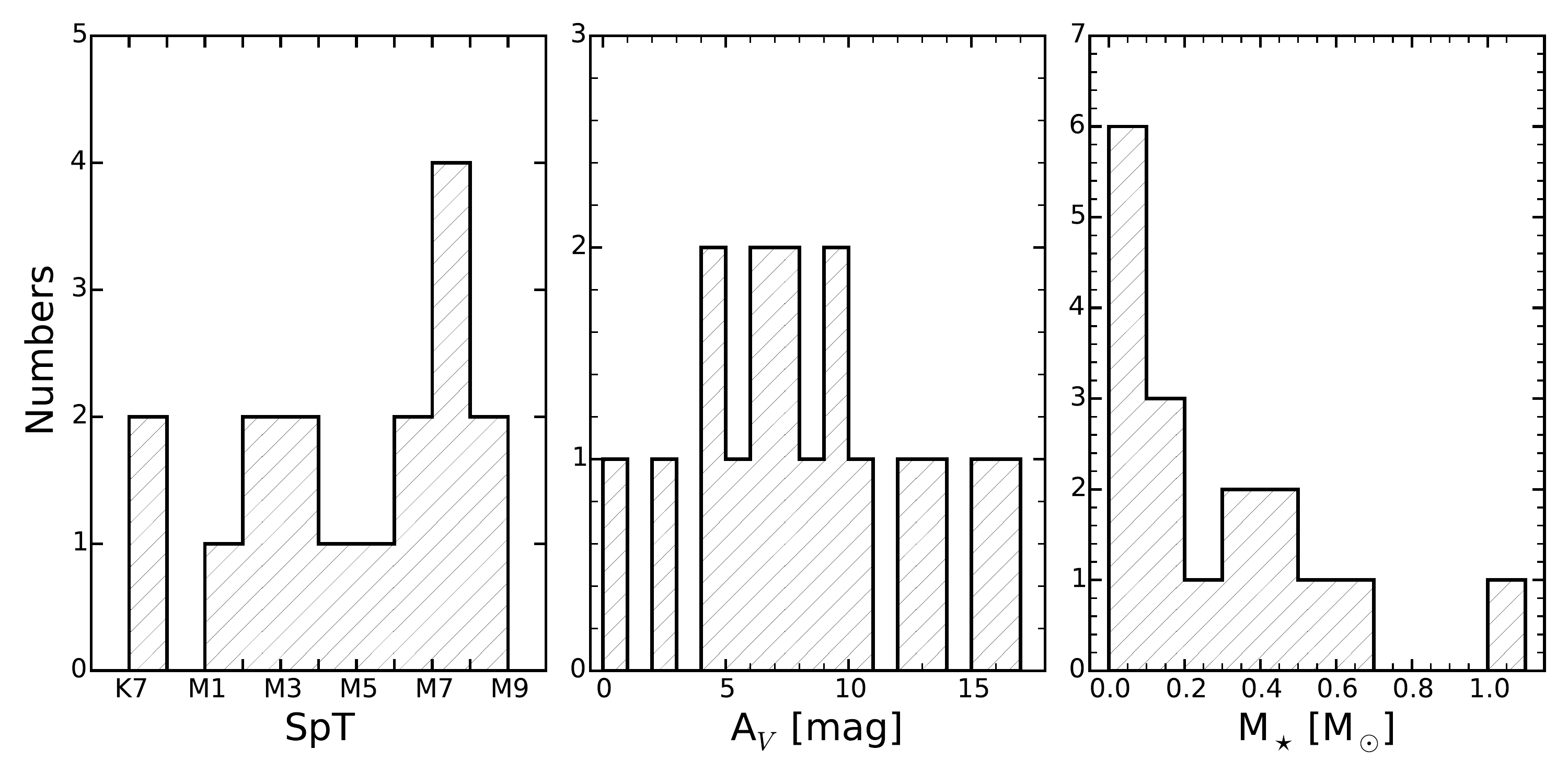}
\caption{Histograms of the properties of the targets. The left-hand panel shows the SpTs, the middle panel shows the $A_V$, and the right-hand panels shows the \mstar. 
According to \mstar \ derived using \citet{Baraffe98} evolutionary tracks, the sample comprises six BDs and two objects with mass just at the hydrogen-burning limit.
     \label{fig::spt_mstar_hist}}
\end{figure}

\setlength{\tabcolsep}{8pt}
\renewcommand{\arraystretch}{1.5}

\begin{table*} 
\centering 
\footnotesize 
\caption{\label{tab::reg::oph_pars}Spectral types, extinction, and physical parameters of the $\rho$-Oph Class~II YSOs} 
\begin{tabular}{lcccccc} 
\hline\hline 
Object & SpT & T$_{\rm eff}$ & $A_V$ & $L_\star$ & $R_\star$ & $M_\star$  \\ 
\hbox{} &  & [K] & [mag] & [L$_\odot$] & [R$_\odot$] & [M$_\odot$] \\ 
\hline 
ISO$-$Oph023 & M7 & 2900$^{+85}_{-50}$ & 9.7$^{+0.2}_{-0.1}$ & 0.040$^{+0.017}_{-0.002}$ & 0.79$\pm0.20$ & 0.07$^{+0.02}_{-0.01}$  \\ 
ISO$-$Oph030 & M7 & 2900$^{+50}_{-50}$ & 4.5$^{+0.2}_{-0.1}$ & 0.067$^{+0.003}_{-0.000}$ & 1.03$\pm0.04$ & 0.10$^{+0.01}_{-0.01}$  \\ 
ISO$-$Oph032 & M6.5 & 2935$^{+50}_{-50}$ & 0.6$^{+0.1}_{-0.1}$ & 0.033$^{+0.000}_{-0.000}$ & 0.70$\pm0.03$ & 0.07$^{+0.01}_{-0.01}$  \\ 
ISO$-$Oph033 & M8 & 2700$^{+150}_{-50}$ & 7.7$^{+0.4}_{-0.2}$ & 0.005$^{+0.001}_{-0.000}$ & 0.32$\pm0.06$ & 0.03$^{+0.02}_{-0.01}$  \\ 
ISO$-$Oph037 & K7 & 4060$^{+50}_{-50}$ & 16.1$^{+0.1}_{-0.2}$ & 0.415$^{+0.000}_{-0.017}$ & 1.31$\pm0.21$ & 1.02$^{+0.01}_{-0.01}$  \\ 
ISO$-$Oph072$^*$ & M3.5 & 3340$^{+50}_{-50}$ & 8.2$^{+0.2}_{-0.1}$ & 0.106$^{+0.005}_{-0.004}$ & 0.97$\pm0.04$ & 0.30$^{+0.01}_{-0.01}$  \\ 
ISO$-$Oph087 & M4.5 & 3200$^{+50}_{-50}$ & 13.3$^{+0.1}_{-0.2}$ & 0.109$^{+0.000}_{-0.005}$ & 1.08$\pm0.05$ & 0.22$^{+0.01}_{-0.01}$  \\ 
ISO$-$Oph094$^\dag$ & M1.5 & 3600$^{+50}_{-235}$ & 10.0$^{+0.8}_{-0.1}$ & 0.009$^{+0.003}_{-0.001}$ & 0.25$\pm0.06$ & 0.40  \\ 
ISO$-$Oph102 & M5 & 3125$^{+50}_{-50}$ & 2.2$^{+0.1}_{-0.1}$ & 0.047$^{+0.000}_{-0.000}$ & 0.74$\pm0.02$ & 0.15$^{+0.01}_{-0.01}$  \\ 
ISO$-$Oph115 & M2 & 3560$^{+50}_{-110}$ & 15.1$^{+0.1}_{-0.7}$ & 0.145$^{+0.006}_{-0.035}$ & 1.00$\pm0.14$ & 0.51$^{+0.01}_{-0.07}$  \\ 
ISO$-$Oph117 & M3.5 & 3340$^{+50}_{-50}$ & 9.1$^{+0.1}_{-0.1}$ & 0.221$^{+0.000}_{-0.000}$ & 1.41$\pm0.04$ & 0.33$^{+0.01}_{-0.01}$  \\ 
ISO$-$Oph123$^*$ & K7$^{**}$ & 4060$^{**}$ & 4.3$^{**}$ & 0.078$^{**}$ & 0.56$^{**}$ & 0.62$^{**}$  \\ 
ISO$-$Oph160 & M7.5 & 2800$^{+50}_{-50}$ & 6.1$^{+0.1}_{-0.2}$ & 0.030$^{+0.000}_{-0.001}$ & 0.73$\pm0.03$ & 0.06$^{+0.01}_{-0.01}$  \\ 
ISO$-$Oph164 & M8 & 2700$^{+50}_{-50}$ & 5.1$^{+0.2}_{-0.1}$ & 0.052$^{+0.002}_{-0.000}$ & 1.05$\pm0.05$ & 0.05  \\ 
ISO$-$Oph165$^\dag$ & M2.5 & 3500$^{+150}_{-135}$ & 12.1$^{+0.6}_{-0.2}$ & 0.034$^{+0.011}_{-0.001}$ & 0.50$\pm0.10$ & 0.40$^{+0.08}_{-0.08}$  \\ ISO$-$Oph176 & M7.5 & 2800$^{+50}_{-50}$ & 6.9$^{+0.1}_{-0.1}$ & 0.057$^{+0.000}_{-0.000}$ & 1.02$\pm0.04$ & 0.06  \\ 
ISO$-$Oph193 & M6 & 3000$^{+50}_{-50}$ & 7.4$^{+0.2}_{-0.2}$ & 0.072$^{+0.003}_{-0.003}$ & 1.00$\pm0.07$ & 0.11$^{+0.01}_{-0.01}$  \\ 
\hline 
\end{tabular} 
\tablefoot{$M_*$ are derived using the evolutionary tracks of \citet{Baraffe98}. $^*$ Object with strong veiling due to accretion. $^{**}$ These stellar parameters are very uncertain and should be used with caution. See Appendix~\ref{app::ind_targ} for a discussion on this target. $^\dag$ Subluminous YSOs. \mstar \ reported with no uncertainties are for objects whose location on the HRD is at the edge of the tabulated evolutionary tracks. }
\end{table*} 

\renewcommand{\arraystretch}{1.0}

\subsection{Derived properties}\label{sect::star_prop}

The stellar parameters derived from the best-fit results are reported in Table~\ref{tab::reg::oph_pars} with their 1$\sigma$ uncertainties. The first two panels of Fig.~\ref{fig::spt_mstar_hist} show the histograms of the SpT and $A_V$ derived for our targets. The vast majority of the YSOs in our sample are of M-type SpT, with 7 out of 17 targets having SpT M6.5 or later. The derived values of $A_V$ are in the range from $\sim$0 mag to $\sim$16 mag, and with a uniform distribution in this interval.  We compare our findings with those in the literature in Sect.~\ref{sect::comp_lit_pars}.

From the best-fit parameters (SpT, $A_V$) we derive $T_{\rm eff}$ using the SpT-$T_{\rm eff}$ relation of \citet{Luhman03}. Stellar luminosity (\lstar) is derived from the ratio of the flux of the reddening corrected observed spectrum to the flux of the best-fit template at $\lambda\sim$1025 nm ($K$) and the known parameters of the template. In particular, when the best-fit template is a Class~III YSO, we know its \lstar \ and distance \citep{Manara13a}, so we can determine \lstar \ for our target simply considering the squared ratio of the distances and the ratio $K$ of the two spectra, as in \citet{Manara13b}. On the other hand, when the best-fit template is a synthetic spectrum, \lstar \ is simply the total flux of the synthetic spectrum rescaled at the distance of $\rho$-Oph and multiplied by $K$, i.e., \lstar=4$\pi\cdot d^2_{\rm \rho-Oph}\cdot K\cdot F_{\rm tot, template}$, which is a similar procedure as in \citet[][and references therein]{Alcala11}. Stellar radii ($R_\star$) are derived from the relation \lstar = 4$\pi R_\star^2 \sigma_B T^4$, while \mstar \ from comparison of the position of the targets on the HRD with the evolutionary tracks by \citet{Baraffe98}. We also show  the distribution of \mstar \ for our targets in the right-hand  panel of
Fig.~\ref{fig::spt_mstar_hist}. According to the parameters derived with the \citet{Baraffe98} evolutionary tracks, six objects in our sample have a mass below the hydrogen-burning limit (\iso023, \iso032, \iso033, \iso160, \iso164, and \iso176), and two more have a value of \mstar \ close to this limit (\mstar$\sim$0.1 \msun, \iso030 and \iso193). The majority of the YSOs have \mstar$\lesssim$0.6 \msun, and only one object has \mstar$\sim$1\msun.

\begin{figure}[]
\centering
\includegraphics[width=0.5\textwidth]{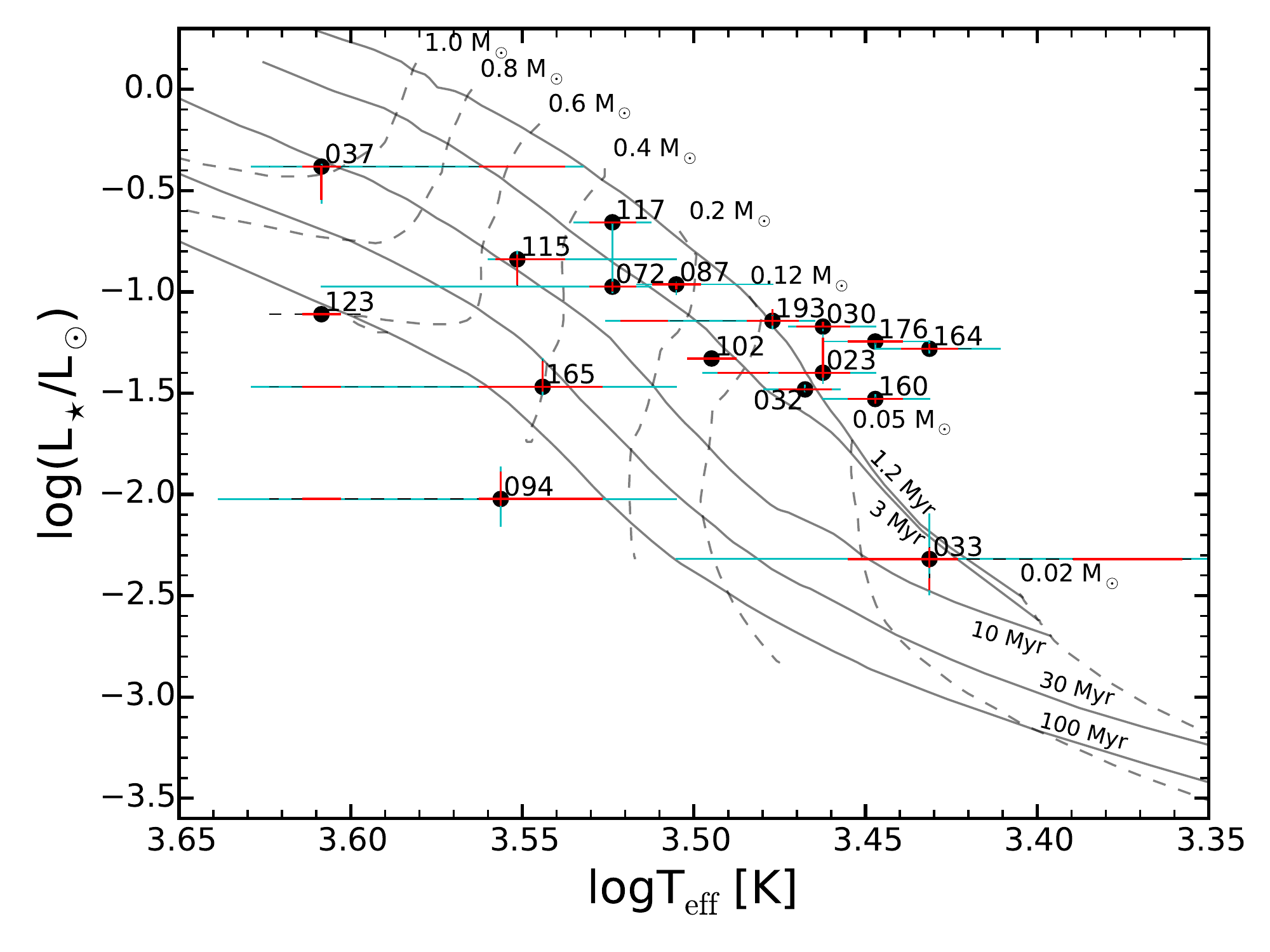}
\includegraphics[width=0.5\textwidth]{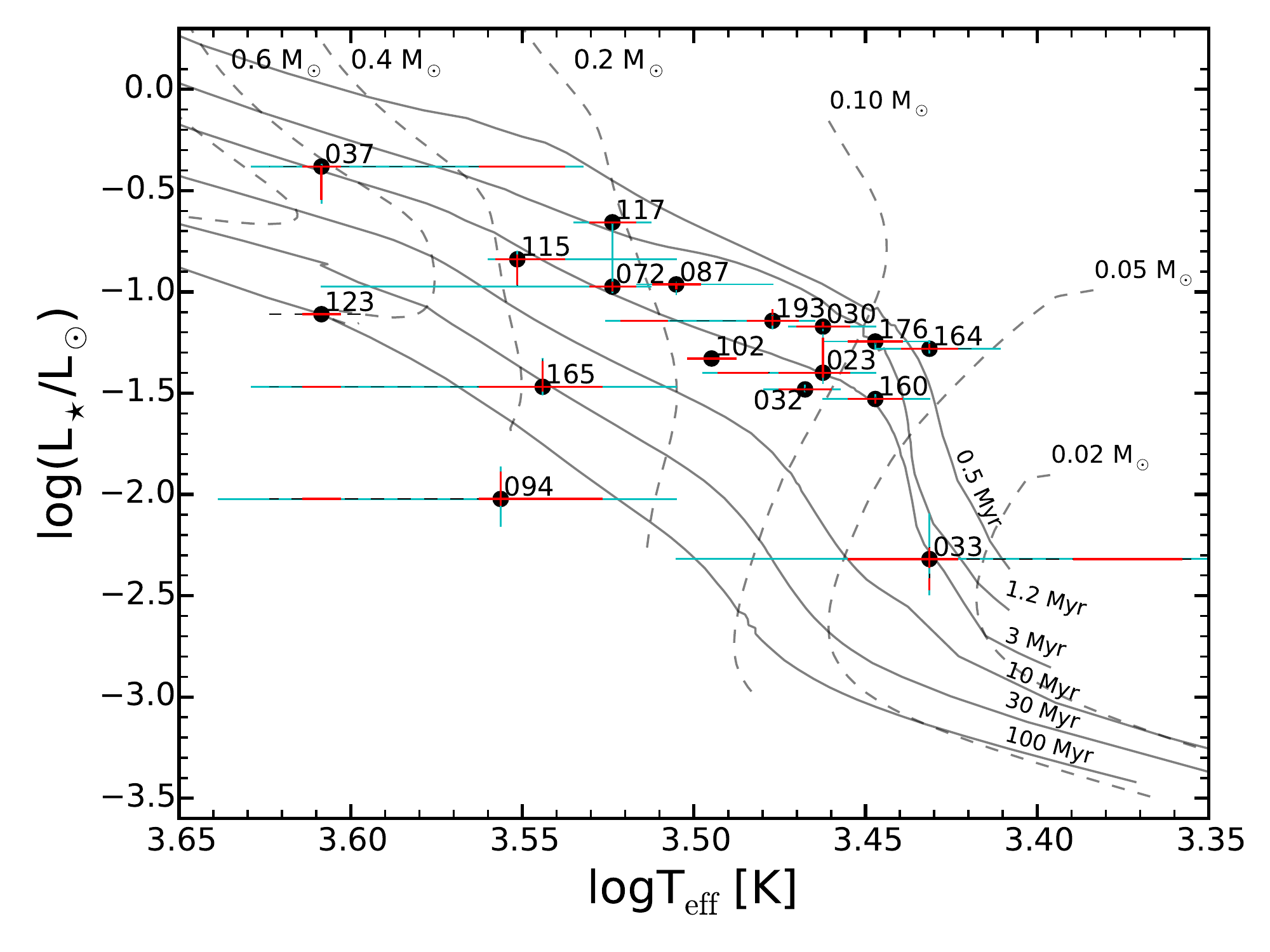}
\caption[Hertzsprung-Russell diagram for the $\rho$-Oph Class~II YSOs analyzed here]{Hertzsprung-Russell diagram for the $\rho$-Oph Class~II YSOs analyzed here. 
The continuous lines show the isochrones, while the dashed lines show the low-mass pre-main sequence evolutionary tracks. The upper panel is obtained using the tracks by \citet{Baraffe98}, while the bottom panel with  tracks by \citet{DAntona}. ISO numbers are reported for all the objects. Red lines represent the 1$\sigma$ error on the parameters, while cyan lines the 3$\sigma$ one. The objects \iso094 and \iso165 are classified as being subluminous probably due to an edge-on disk for the former and to a still partly optically thick envelope in the latter, which is a Class~I target. The stellar parameters and the position on the HRD of \iso123 is very uncertain due to the very strong veiling due to accretion, which makes the spectral classification extremely difficult.
     \label{reg::fig::hrd_oph}}
\end{figure}

The position of the targets on the HRD is shown in Fig.~\ref{reg::fig::hrd_oph}, together with the 1$\sigma$ (red lines) and 3$\sigma$ uncertainties (blue lines) on the estimated parameters. An additional $\lesssim$0.2 dex uncertainty on \lstar \ should be considered, as this is the error on the luminosity of the templates due to uncertainties in the distance and in the flux calibration of the spectra \citep[e.g.,][]{Manara13a,Alcala14}. The HRD is shown with two different sets of evolutionary tracks, those by \citet[][upper panel]{Baraffe98} and those by \citet[][bottom panel]{DAntona}. The majority of the targets (14 out of 17) is found to be located on the HRD at positions compatible with young ($<$5-10 Myr) ages according to both sets of models, and at ages $\lesssim$3 Myr according to \citet{DAntona} evolutionary tracks. The latter set of models appears to reproduce better the positions on the HRD of the BDs in the sample, in particular, of ISO$-$Oph176 and ISO$-$Oph164. We however adopt the parameters derived using the evolutionary tracks from \citet{Baraffe98} for consistency with previous studies of accretion in nearby star-forming regions \citep[e.g.,][]{Alcala14}. 

For three objects (\iso123, \iso165, and \iso094) the position on the HRD is significantly closer to the main sequence, thus suggesting an older age than the rest of the targets. However, none of these targets are probably really old. The first object, \iso123, is extremely veiled and this makes it difficult to detect the photospheric continuum. Therefore, the stellar parameters for this target are very uncertain. We discuss this case in detail in Appendix~\ref{app::ind_targ}. Regarding ISO$-$Oph165, mid-infrared data have shown that this object is still embedded in its own parental envelope, and it is classified as a Class~I YSO \citep{McClure10}. For this reason, the stellar parameters are uncertain, $A_V$ is high, and an erroneously lower stellar luminosity estimate is possible. Finally, we suggest that \iso094 is a target whose circumstellar disk is edge-on, as its spectrum has an almost undetected continuum, also with  a very low SNR in the NIR arm. We note that distance uncertainties for these objects are probably not leading to an erroneous location on the HRD, as a factor of $\sim$2.5 larger distance would be needed to position \iso165 closer to the bulk of the population, and this factor is even higher ($\sim$5) for \iso094. For the remainder of this work, we  refer to ISO$-$Oph165 and ISO$-$Oph094 as subluminous YSOs and use a different symbol in the plots to differentiate these from the other targets. Additional information and discussion on these objects is given in Appendix~\ref{app::ind_targ}.

\subsection{Comparison with previous results}\label{sect::comp_lit_pars}
Most of the targets analyzed here have been previously observed in near-infrared spectroscopic or photometric studies \citep[e.g.,][]{Wilking99,Luhman99,Natta02,Natta04,Natta06} or with optical spectroscopy \citep[e.g.,][]{Wilking05}. The stellar parameters (SpT, $A_V$) derived in these works for our targets are reported in Table~\ref{tab::lit} and in Table~\ref{tab::natta_125} for \citetalias{Natta06}. 

\begin{figure}[]
\centering
\includegraphics[width=0.5\textwidth]{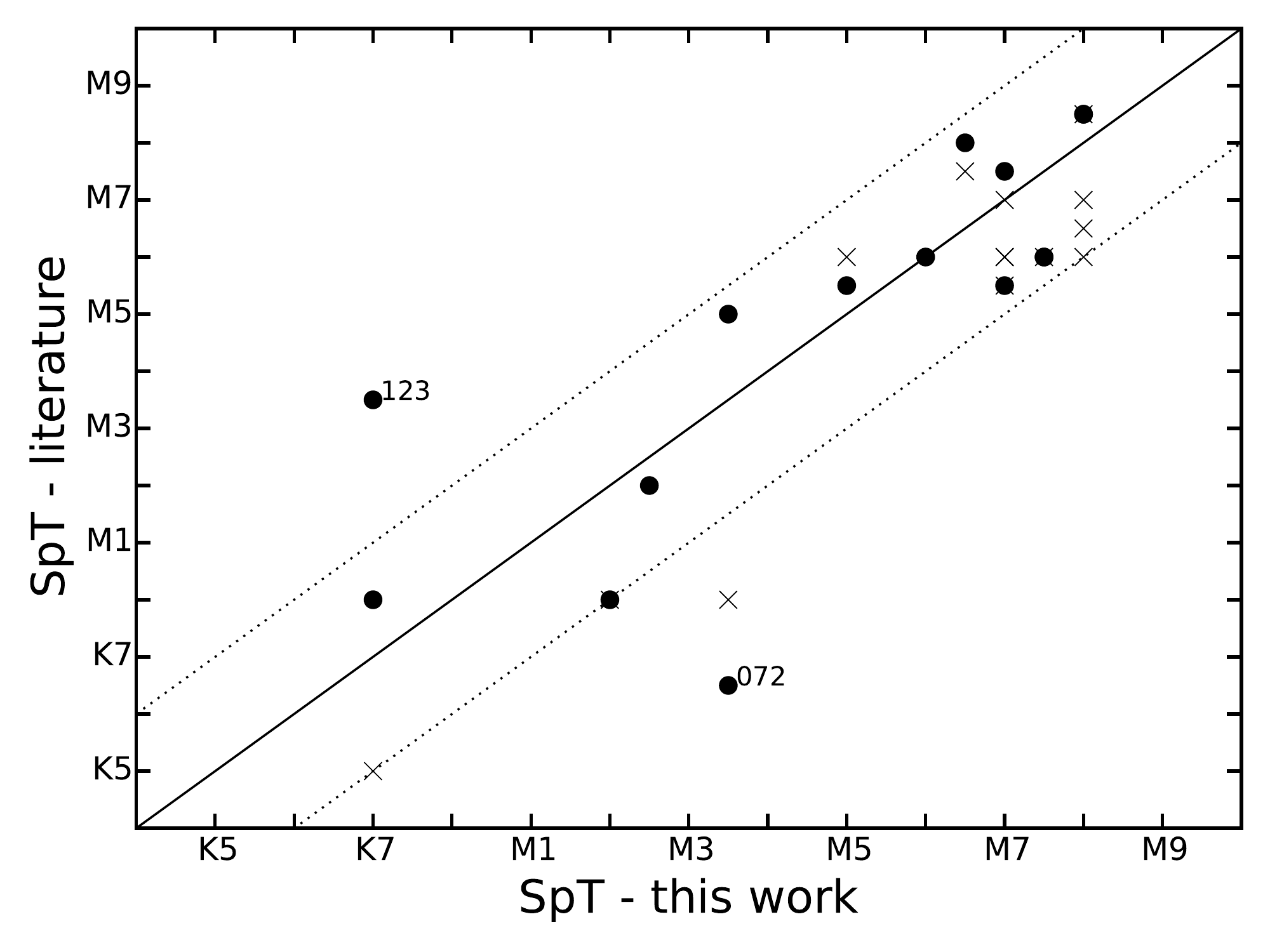}
\caption{Comparison of the SpT derived here for the targets (Sect.~\ref{sect::star_prop}) with those reported in the literature (see Table~\ref{tab::lit} for the references). Black circles report the most recent SpT estimate found in the literature, while black crosses are used for other SpT estimates available. The dotted lines show the $\pm$2 spectral subclasses difference between the two values. The larger discrepancies are for \iso123, classified here as K7 except with very uncertain parameters, and \iso072, which we classify as M3.5, and was reported to be K6.5 in the literature.
     \label{fig::spt_comp}}
\end{figure}

The comparison between the SpTs derived here and those obtained with spectroscopy in the literature (Table~\ref{tab::lit}) is shown in Fig.~\ref{fig::spt_comp}, where all the literature estimates are reported using circles for the most recent one and crosses for the older results. All our SpT estimates agree within 1 or 2 spectral subclasses with previous studies, with the only two exceptions being \iso123, which was classified as M3.5, while we find a very uncertain best fit with SpT K7, and \iso072, which we found to have a SpT M3.5, while it was previously classified as K6.5 by \citet{Wilking05}. In both cases, the targets are strongly veiled, and thus their spectral type classification is subject to large uncertainties that can justify these differences.  
Finally, we derive for \iso117 a SpT of M3.5, which is in better agreement with the SpT M5 reported by \citet{McClure10} than with the SpT K8 found by \citet{Gatti06}. 

According to \citetalias{Natta06}, only four objects in our sample should have \mstar \ well above the hydrogen burning limit (\iso037, \iso072, \iso087, and \iso117). The result of our analysis confirms that all these YSOs are VLMS. At the same time, among the other 13 targets that were classified as BDs in \citetalias{Natta06} only 8 are confirmed as BDs by our analysis, if we also consider \iso030 and \iso193 as BDs. Three more targets that were classified as BDs are those with highly uncertain stellar parameters, as they are either subluminous (\iso094, \iso165) or strongly veiled (\iso123).  Regarding the remaining two objects that were thought to be BDs, we classify \iso102 as of M5 SpT and with a \mstar \ slightly above the hydrogen-burning limit (\mstar = 0.15 \msun), and this estimate is also compatible  with \citet{Wilking05}. Finally, our derived \mstar \ for \iso115 is 0.5 \msun, while it was reported to be just above the hydrogen burning limit by \citetalias{Natta06}. However, this object was reported to have a SpT M0 by other authors \citep{Gatti06,McClure10}, which confirms it is a low-mass YSO, not a BD.

\begin{figure}[]
\centering
\includegraphics[width=0.5\textwidth]{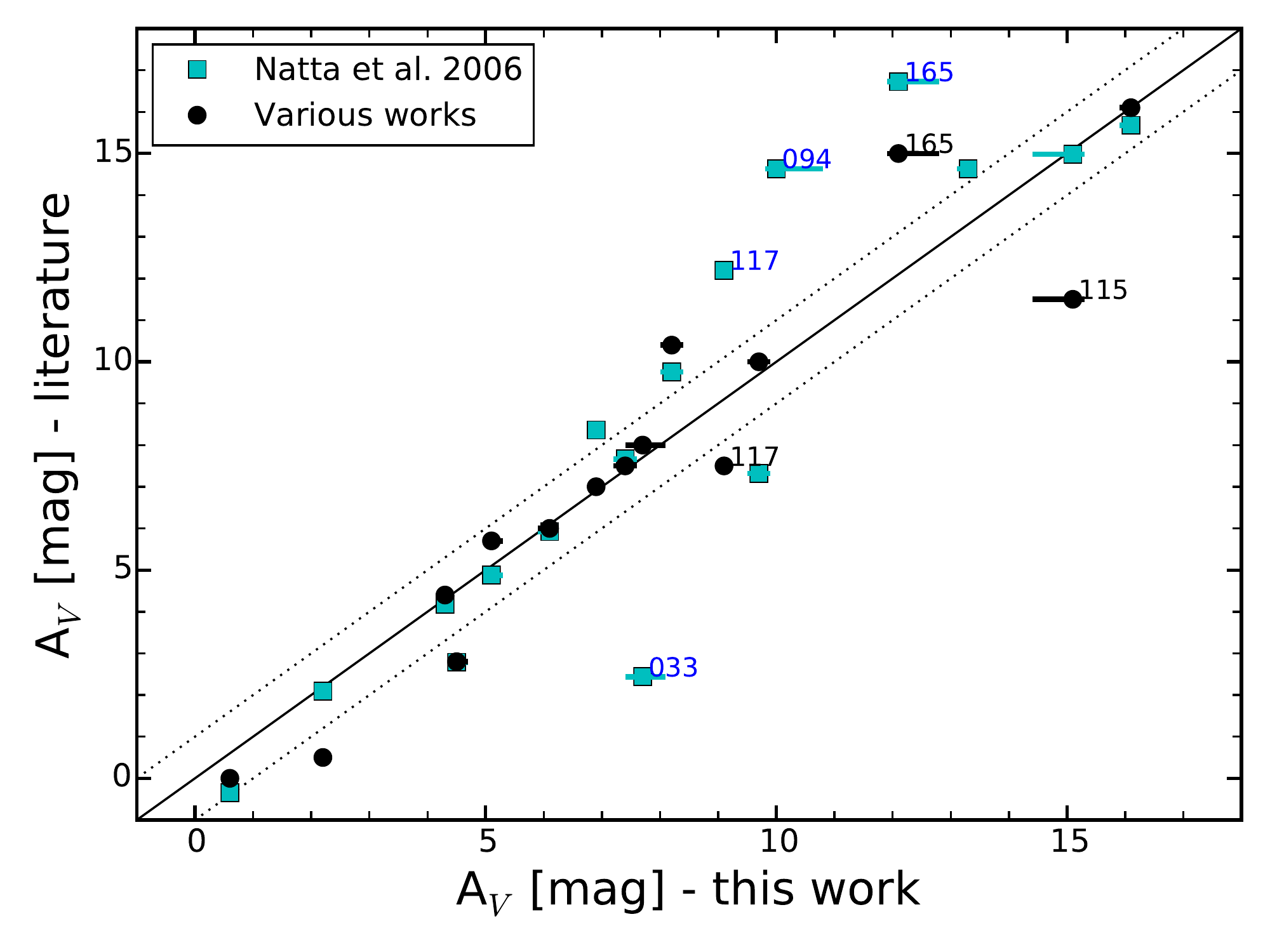}
\caption{Comparison of the $A_V$ derived here for the targets (Sect.~\ref{sect::star_prop}) with those reported in the literature. The latter are shown with \textit{cyan squares} in the case of values derived by \citetalias{Natta06} and in \textit{black circles} when from other works (see Table~\ref{tab::lit} for the references). The dotted lines show the $\pm$1 mag difference between the two values. One sigma uncertainties in our estimates are reported, and are sometimes smaller than the symbol. 
     \label{fig::av_comp}}
\end{figure}

In Fig.~\ref{fig::av_comp} we show  the comparison between the values of $A_V$ derived here with those derived by \citetalias{Natta06} for all the objects in our sample (\textit{cyan squares}) and by other works for most of the targets (\textit{black circles}). We see that most values agree within a typical difference up to $\sim$1 mag. The largest differences are found for \iso165, where both \citetalias{Natta06} and \citet{McClure10} report a value of $A_V\sim$15-17 mag higher than our estimate $A_V$=12.1 mag, and then for \iso094, reported to have $A_V$=14.6 mag by \citetalias{Natta06} and found to have $A_V$=10 mag here. In one case the derived value differs from the estimates of \citetalias{Natta06} but not with other studies (\iso033) or viceversa (\iso115), while the differences of $A_V$ derived for \iso117 are higher than 2 mag with all previous studies, but in different directions.

Finally, we compare the \lstar \ derived here with those determined by \citetalias{Natta06} and modified to account for the correct distance estimate (see Table~\ref{tab::natta_125}). This is a very interesting comparison, as \lstar \ is derived with two independent methods. Therefore, it is very important to note that the two methods lead to similar results for all targets, with differences smaller than a factor $\sim$2. The only exception is \iso033, but we note that the revised analysis by \citet{Comeron10} leads to the same value of \lstar \ as the value we derived here. 

To summarize, we find, in most cases, values for $A_V$ and \lstar \ to be compatible with those by \citetalias{Natta06} and other literature studies, but that the derived \mstar \ differ from the literature estimate. The main reason for the different estimates of \mstar \ is the different T$_{\rm eff}$ determined here with a more robust method than those by \citetalias{Natta06} and with better data than most previous spectroscopic studies.

\begin{figure}[]
\centering
\includegraphics[width=0.5\textwidth]{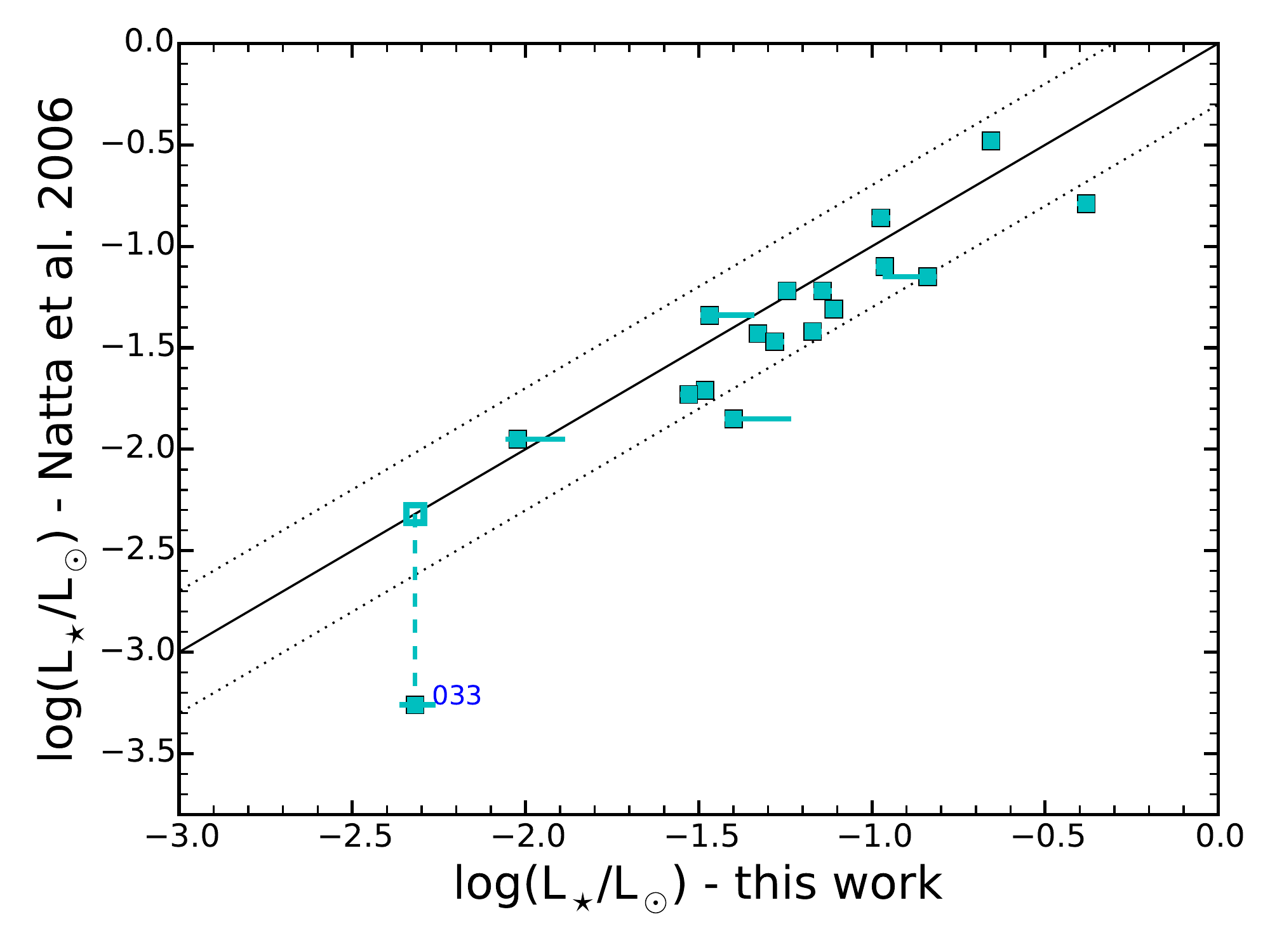}
\caption{Comparison of the \lstar \ derived here for the targets (Sect.~\ref{sect::star_prop}) with those from \citetalias{Natta06} and corrected for distance as discussed by \citet{Rigliaco11a}. The dotted lines show a difference between the \lstar \ values of a factor 2. The only object with a significantly larger difference is ISO$-$Oph033, but its \lstar \ has been revised by \citet{Comeron10} and the latter value, in agreement with ours, is reported with an empty symbol. One sigma uncertainties in our estimates are reported, and are sometimes smaller than the symbol.
     \label{fig::lstar_comp}}
\end{figure}

\subsection{Spectral features}
We visually investigate the spectra with SNR$>$10 in the continuum at $\sim$700 nm for the presence of the Li~I absorption line at $\lambda$670.78 nm, a known proxy of youth in YSOs. We clearly detect this line in the spectrum of ISO$-$Oph032, while it is only tentatively detected in the spectra of ISO$-$Oph030, ISO$-$Oph102, and ISO$-$Oph123. In the latter, the SNR of the spectrum is very high, but the strong veiling makes all the absorption lines very weak. For all these targets, we report  the presence of lithium for the first time, while the low SNR of the other spectra hinder the detection of this line in the rest of the sample. 

Various permitted emission lines are detected in the spectra of our targets, and these are used in the next section to derive the accretion luminosity of the targets. In few spectra with high SNR, we also detect forbidden emission lines. In particular, the [OI] $\lambda$ 6300 \AA \ line is detected in the spectra of ISO$-$Oph030, ISO$-$Oph032, ISO$-$Oph102, and ISO$-$Oph123. The analysis of these lines is out of the scope of this paper and will be discussed in a following work.

\begin{figure}[h!]
\centering
\includegraphics[width = 0.5\textwidth]{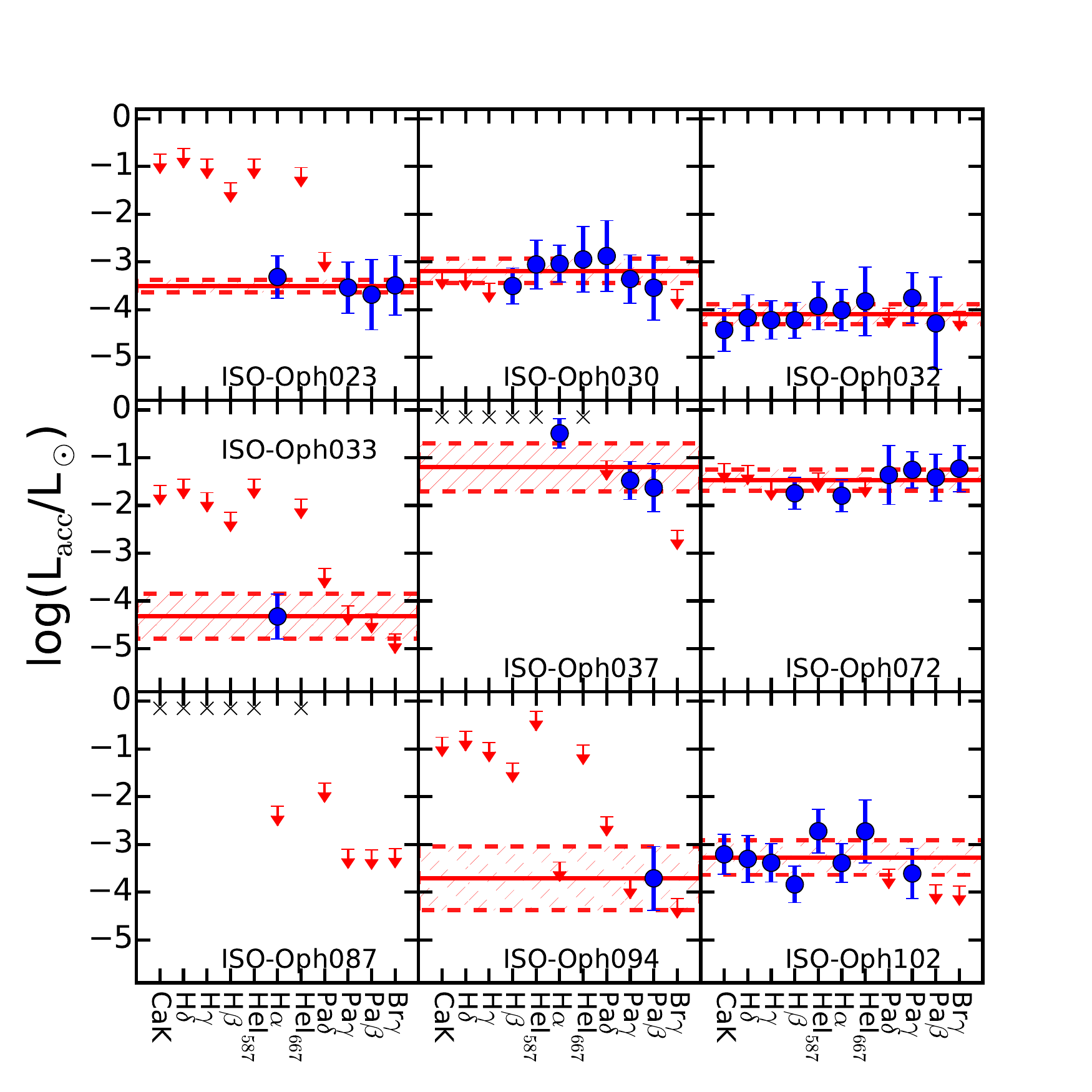}
\includegraphics[width = 0.5\textwidth]{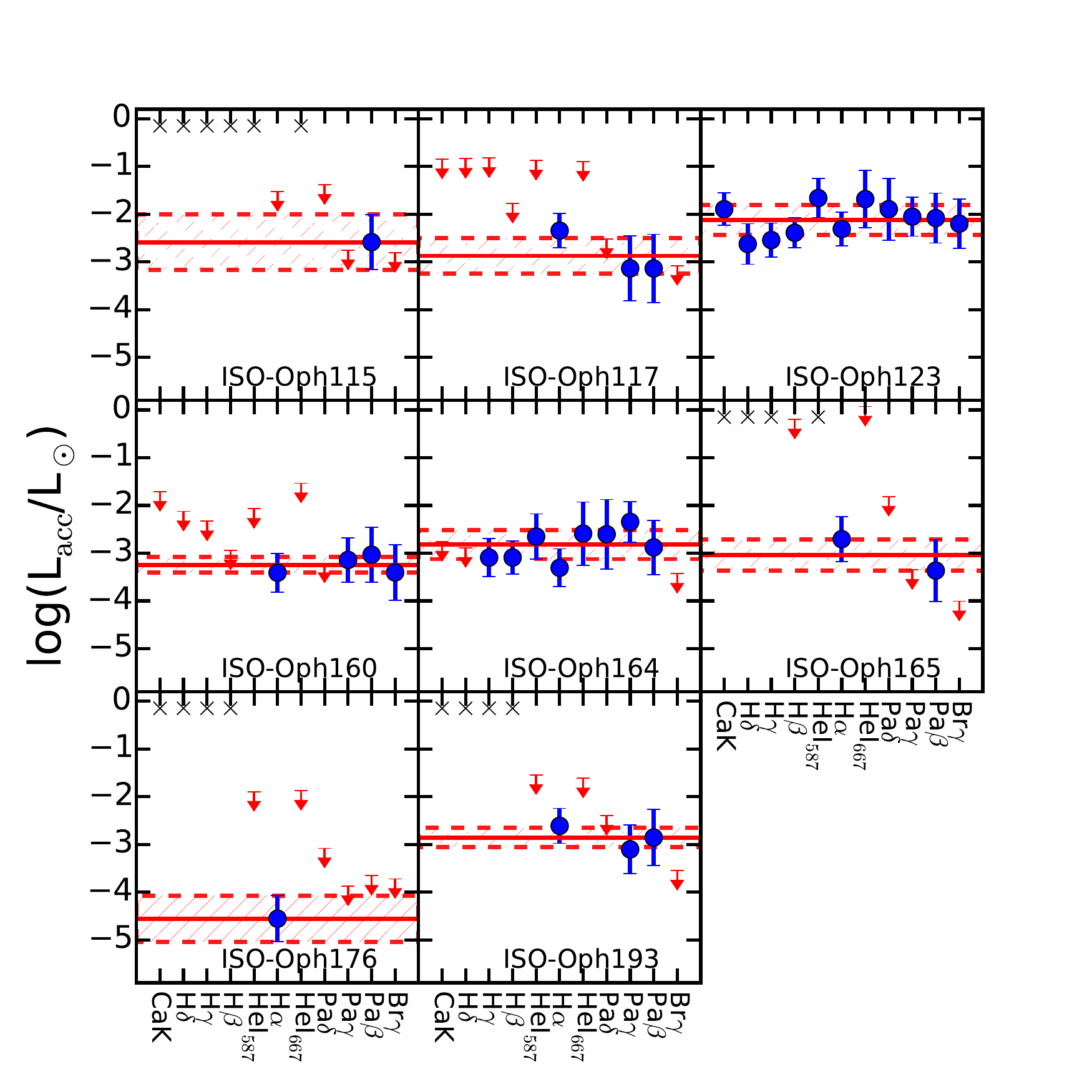}
\caption{Accretion luminosity derived from various emission lines luminosity for the $\rho$-Oph targets. Each subplot shows the value of $\log$(\lacc/\lsun) \ derived using the various indicators reported on the x-axis (CaK, H$\delta$, H$\gamma$, H$\beta$, He$_{\lambda 587 {\rm nm}}$, H$\alpha$, He$_{\lambda 667 {\rm nm}}$, Pa$\delta$, Pa$\gamma$, Pa$\beta$, Br$\gamma$) and in order of increasing wavelength. The red solid line is the average values obtained from the detected lines, while the dashed lines are the 1$\sigma$ standard deviation of this value. Black crosses are for upper limits out of the range of the plot, i.e., larger than log(\lacc/\lsun)$\sim$-0.1. As upper limits depend on the rms noise of the spectra, they are significantly higher than measurements when the SNR is low, in particular, in the UVB arm. 
     \label{reg::fig::lline_oph}}
\end{figure}

\section{Accretion properties of $\rho$-Ophiucus young stellar objects}\label{sect::acc_prop}

The accretion luminosity for our targets is derived using relations between the luminosity of emission lines (\lline) and \lacc , which have been calibrated by \citet{Alcala14}. These represent the only suitable way to determine \lacc \ from our spectra, as the UV-excess is not detected in the spectra because of the high extinction of the region. This is an indirect method that is less accurate than direct fitting of the UV-excess, but it has been shown that \lacc \ determined in this way are consistent with direct measurements of \lacc \ when multiple lines are used \citep[e.g.,][]{Rigliaco12,Alcala14}. 

For this analysis, we then select 11 emission lines that are usually bright and whose \lline \ has a good correlation with \lacc, namely CaK, H$\delta$, H$\gamma$, H$\beta$, He$_{\lambda 587 {\rm nm}}$, H$\alpha$, He$_{\lambda 667 {\rm nm}}$, Pa$\delta$, Pa$\gamma$, Pa$\beta$, Br$\gamma$. We determine the flux of the emission lines from the flux-calibrated and extinction-corrected spectra using an automatic Python procedure that determines the value of the continuum and the extent of the line, which is then directly integrated with no additional modeling. All the line extent determination of the automatic procedures are checked by eye, and the derived flux of the lines are compatible with results obtained using different methods, such as direct integration using the {\em splot} package under IRAF. The error on the line flux is obtained propagating the 1$\sigma$ standard deviation on the continuum flux over the integration window. For nondetected lines, we calculate the 3$\sigma$ upper limits with the relationship ${\rm 3 \times F_{\rm noise} }\times \Delta\lambda$, where F$_{\rm noise}$  is the rms flux-noise in the region of the line and $\Delta\lambda$ is the expected average line width, assumed to be 0.2\,nm. The fluxes and equivalent width (EW) of the lines, together with their respective errors, are reported in Table~\ref{tab::perm_line_flux}.

All the objects analyzed here with detected H$\alpha$ line have EW$_{\rm H\alpha}$ well above the threshold for accretors reported by \citet{WB03}, and thus are confirmed accretors. This line is not detected in the spectra of \iso087, \iso094, and \iso115, which have SNR$\sim$0 on the continuum adjacent to the line. For the latter target, we are able to determine \lacc \ from the Pa$\beta$ line and this leads to a value above the chromospheric noise level, i.e., the intensity of the chromospheric emission in a nonaccreting YSO, determined by \citet{Manara13a} for objects with this T$_{\rm eff}$. The value of \lacc \ determined for \iso094 from the luminosity of the Pa$\beta$ line seems compatible with pure chromospheric emission. However, this object is probably observed edge-on, thus the measured line intensity is probably only a lower limit of the real line intensity. Finally, no emission lines are detected for \iso087, which could then possibly be a nonaccreting YSO.

\begin{table} 
\centering 
\footnotesize 
\caption{\label{tab::reg::oph_acc}Accretion luminosity and mass accretion rates of the $\rho$-Oph YSOs} 
\begin{tabular}{lccl} 
\hline\hline 
Object & $\log L_{\rm acc}$ & $\log \dot{M}_{\rm acc}$  & Detected lines \\ 
\hbox{} &  [L$_\odot$] & [M$_\odot$/yr] & [\#] \\ 
\hline 
ISO$-$Oph023 & -3.51 & -9.85 & 4\\ 
ISO$-$Oph030 & -3.19 & -9.58 & 7\\ 
ISO$-$Oph032 & -4.09 & -10.49 & 9\\ 
ISO$-$Oph033 & -4.33 & -10.70 & 1\\ 
ISO$-$Oph037 & -1.20 & -8.49 & 3\\ 
ISO$-$Oph072 & -1.47 & -8.36 & 8\\ 
ISO$-$Oph087 & $<$-2.21 & $<$-8.92 & 0 \\ 
ISO$-$Oph094 & -3.71 & -11.32 & 1\\ 
ISO$-$Oph102 & -3.27 & -9.98 & 8\\ 
ISO$-$Oph115 & -2.58 & -9.69 & 1\\ 
ISO$-$Oph117 & -2.87 & -9.64 & 3\\ 
ISO$-$Oph123 & -2.12 & -9.56 & 11\\ 
ISO$-$Oph160 & -3.25 & -9.56 & 4\\ 
ISO$-$Oph164 & -2.82 & -8.90 & 8\\ 
ISO$-$Oph165 & -3.04 & -10.34 & 2\\ 
ISO$-$Oph176 & -4.60 & -10.73 & 1\\ 
ISO$-$Oph193 & -2.86 & -9.30 & 3\\ 
\hline 
\end{tabular} 
\tablefoot{\lacc \ is obtained as the mean of the values from the different emission lines. \macc \ are derived using \mstar \ determined from the evolutionary tracks of \citet{Baraffe98}. The last column reports the number of emission lines detected and used to derive \lacc. } 
\end{table} 

The luminosity of the lines is calculated as \Ll ~$ = 4 \pi  d^2 \cdot f_{\rm line}$, where $d$ is the distance of $\rho$-Oph and $f_{\rm line}$ the flux of the line. The value of \lacc \ is then determined as the average of the values of \lacc \ determined from the \lacc-\lline \ relations from the detected lines, while the error is the standard deviation of these values. Figure~\ref{reg::fig::lline_oph} 
shows the values of \lacc \ determined with each line for all the targets. The accretion luminosity for the targets are reported in the second column of Table~\ref{tab::reg::oph_acc}, while the number of detected emission lines used to calculate \lacc \ is available in the last column of the same table. We estimate an upper limit on \lacc \ for \iso187 from the upper limit on the H$\alpha$ line and use this value in the analysis.

Finally, \macc \ is determined from the values of \lacc \ just obtained and from the stellar parameters $M_\star$ and $R_\star$ determined as explained in the previous section. These parameters are related by the relation \macc =  1.25 $\cdot$ \lacc $R_\star /(G M_\star )$ \citep[e.g.,][]{Hartmann98}, which assumes accretion from a radius $R_m=5R_\star$. The final value is reported for every object in the third column of Table~\ref{tab::reg::oph_acc}.

\subsection{Comparison with previous results}

The most comprehensive catalog of accretion rates in $\rho$-Oph was compiled by \citetalias{Natta06}. Here we compare the accretion properties we derived for our sample with the accretion properties compiled by N06 for the same objects. As already mentioned in Sect.~\ref{sect::comp_lit_pars}, the results may differ in particular on the estimates of T$_{\rm eff}$ and \mstar \ due to the different methodologies. Indeed, \macc \ depends on various stellar parameters (\mstar, $R_\star$, thus \lstar \ and $T_{\rm eff}$), which  have been determined here with a more detailed analysis. As we reported in Sect.~\ref{sect::intro}, \citetalias{Natta06} did not have estimates for the SpT for most of the targets and had to assume that they were located on the HRD on a single isochrone after determining \lstar \ from the infrared colors. Our estimates of \lstar \ are consistent with those of N06, except that the T$_{\rm eff}$ estimated by \citetalias{Natta06} may differ substantially from ours in many cases, up to the extreme cases of the subluminous objects, which could not be classified as such by \citetalias{Natta06} given their methodological assumption. This results in different \mstar \ and $R_\star$ between the two works. Moreover, \citetalias{Natta06} use the intensity of a single emission line to derive \lacc, while we measure this quantity using multiple, up to 11, emission lines. We are thus less affected by scatter in one single \lacc-\lline \ relation or by other processes that modify the intensity of a single line.

\begin{figure}[]
\centering
\includegraphics[width=0.48\textwidth]{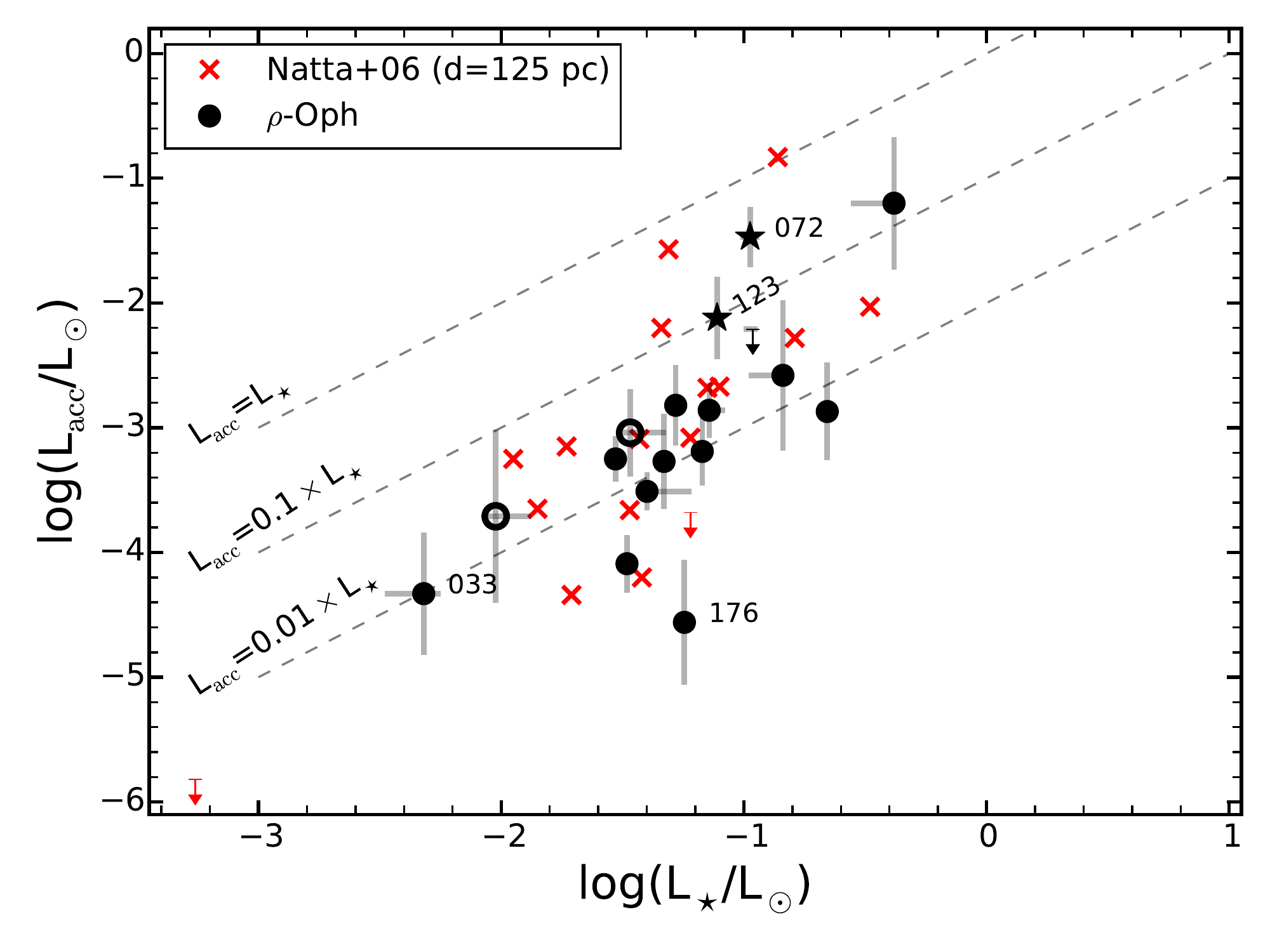}
\caption{Comparison of accretion rate luminosity as a function of stellar luminosity for our sample with data from \citetalias{Natta06} for the same targets. Values from \citetalias{Natta06} are shown with \textup{\textup{\textit{\textup{red crosses}}}}, while our results with black circles. Empty black circles are used for subluminous targets and downward arrows for upper limits. 
Dashed lines are for different \lacc/\lstar \ ratios, going downward from 1, to 0.1, to 0.01, as labeled. The object most to the left-hand side of the plot in both samples is \iso033.
     \label{fig::oph_lacc_lstar_both}}
\end{figure}

We first show in Fig.~\ref{fig::oph_lacc_lstar_both} the comparison of the \lacc-\lstar \ relation for the objects included in our sample using our own results (black points) and those from \citetalias{Natta06} (red symbols). The two distributions appear very similar, with the majority of the points located between the \lacc=0.1 \lstar \ and \lacc=0.01 \lstar \ lines. In general, most of the targets are located very close to the \lacc=0.01 \lstar \ line, with slightly smaller dispersion than in the previous work reported here. This is also the case  for the two subluminous targets, which appear on this plot in the same position as the rest of the YSOs. The value of \lstar \ we determine here for \iso033 (bottom left point in both samples) is larger than that by \citetalias{Natta06}, as discussed earlier, leading to a position of this target on the \lacc-\lstar \ plane closer to the bulk of the population than before. The object with the lower  \lacc/\lstar \ ratio is \iso176, but its \lacc \ is still compatible with a line intensity being genuinely due to accretion, and not to chromospheric emission. 
Finally, the position of \iso123 on the \lacc=0.1 \lstar \ line is probably due to the already discussed uncertainties in its parameters. In reality, because of its strong emission, both in lines and continuum, this object is expected to fall closer to the \lacc=\lstar \ line in this diagram, as it was according to \citetalias{Natta06}.  Appendix~\ref{app::ind_targ} provides further discussion on this target.

\begin{figure}[]
\centering
\includegraphics[width=0.48\textwidth]{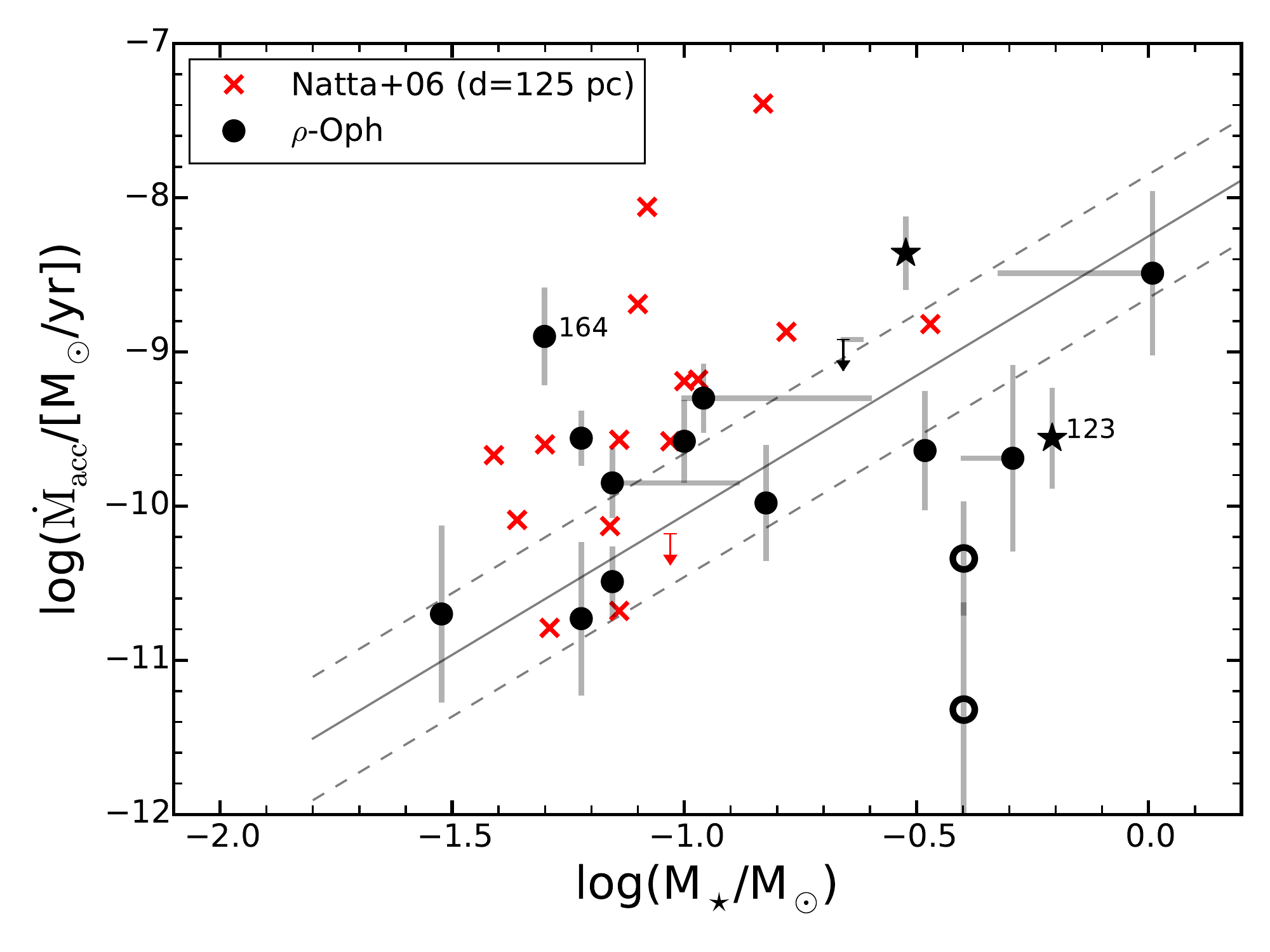}
\caption{Mass accretion rate as a function of mass for the $\rho$-Oph sample. Values derived for the objects analyzed here are reported with \textit{\textup{black}} markers, 
values from \citetalias{Natta06} are shown with red symbols. Empty black circles are used for subluminous targets and downward arrows are used for upper limits. The continuous line represents the linear fit of this relation by \citet{Alcala14} for a sample of accreting objects in Lupus. The dashed lines represent the 1\,$\sigma$ deviation from the fit. 
     \label{fig::oph_Macc_Mstar_both}}
\end{figure}

The comparison of our measurements of \macc \ with those by \citetalias{Natta06} is shown in Fig.~\ref{fig::oph_Macc_Mstar_both}, where we plot the $\log$\macc \ vs $\log$\mstar \ values for the objects analyzed here, again using  the black symbols for our results and red symbols for those by \citetalias{Natta06}. With respect to the results of \citetalias{Natta06}, the largest discrepancies are in the derived \mstar, which in many cases are larger  in our work. This is a consequence of the later SpT determined for various objects with respect to those assumed by \citetalias{Natta06} (see discussion in Sect.~\ref{sect::comp_lit_pars}). At the same time, the range of values of \macc \ is similar for the two works, but our estimates lead to a smaller number of strongly accreting BDs than those of \citetalias{Natta06}. In particular, all the objects in our sample with reliable (sub)stellar and accretion parameters are found on the \macc-\mstar \ to follow the best-fit relation found by \citet[][shown here with a solid gray line within its dispersion, indicated with dashed lines]{Alcala14}, while $\sim$50\% of these targets were above this region for \citetalias{Natta06}. The only object in our sample significantly above the dispersion of the best-fit relation by \citet{Alcala14} is \iso164. For this target, \mstar \ remains similar to that derived by \citetalias{Natta06}, while \macc \ increases by more than an order of magnitude. This object appears to have a \lstar \ slightly higher than the rest of the objects analyzed here with similar SpT (see Fig.~\ref{reg::fig::hrd_oph}), and this could lead to a slightly higher \macc. The only targets that are located on the \macc-\mstar \ plane significantly below the bulk of the population are the two subluminous targets. 

We can thus conclude from this analysis that, in our limited sample of BDs and VLMS, we see a smaller number of heavily accreting targets than \citetalias{Natta06}. This strongly suggests that BDs and VLMS in $\rho-$Oph accrete at a rate similar to those in other regions, as we discuss in the next section, contrary to the claim of \citetalias{Natta06}. As the typical \lacc/\lstar \ ratio in our targets is similar to that determined by \citetalias{Natta06}, our suggestion is that the differences in the \macc-\mstar \ relation is mostly because of the different (sub)stellar parameters determined here. Given that we do not have assumptions about the positions of the targets on the HRD, and we have larger wavelength coverage that lead to a better constraint of $A_V$, SpT, and \lstar, we are confident that our results are more reliable and that the distribution we observe in the \macc-\mstar plane for our sample resembles the real distribution. A larger study on the whole sample of \citetalias{Natta06} with a data set similar to ours would further reinforce our finding. At present we do not attempt to fit our results as we have few objects sparsely sampled in the parameter space (\lstar, \lacc, \mstar, \macc).

\section{Accretion in $\rho$-Oph compared with nearby star-forming regions}\label{sect::acc_comp}

As discussed in Sect.~\ref{sect::intro}, the main interests in the data set presented here are to determine whether BDs in $\rho-$Oph have significantly higher accretion rates than in other star-forming regions, and to study a very young region to compare the accretion properties of its targets with those of targets located in older star-forming regions. We have just discussed how our results differ from those of \citetalias{Natta06} in the \macc-\mstar \ plane, leading to lower accretion rates for BDs than those reported by them. Here we compare our findings with those of 
\citet{Alcala14} in the Lupus I and III clouds, and those by \citet{Rigliaco12} in the $\sigma$-Orionis ($\sigma$-Ori) cluster. These are selected as comparison sample as they include a significant number of VLMS and some ($\sim$5) BDs, and have been studied using spectra obtained with the same instrument. In addition, our accretion rates are determined using the \lacc-\lline \ relation determined from these samples, hence the biases in the \macc \ estimates are minimized than when comparing \macc \ estimates with other data sets. As these data have been collected as part of the Italian guaranteed time observation \citep[GTO;][]{Alcala11}, in the following we refer to the whole comparison sample as GTO sample. Our targets are also interesting as they enlarge the number of targets at low \mstar, in particular, there are six more BDs and two more objects with \mstar \ just at the hydrogen burning limit.

\begin{figure}[!t]
\centering
\includegraphics[width=0.45\textwidth]{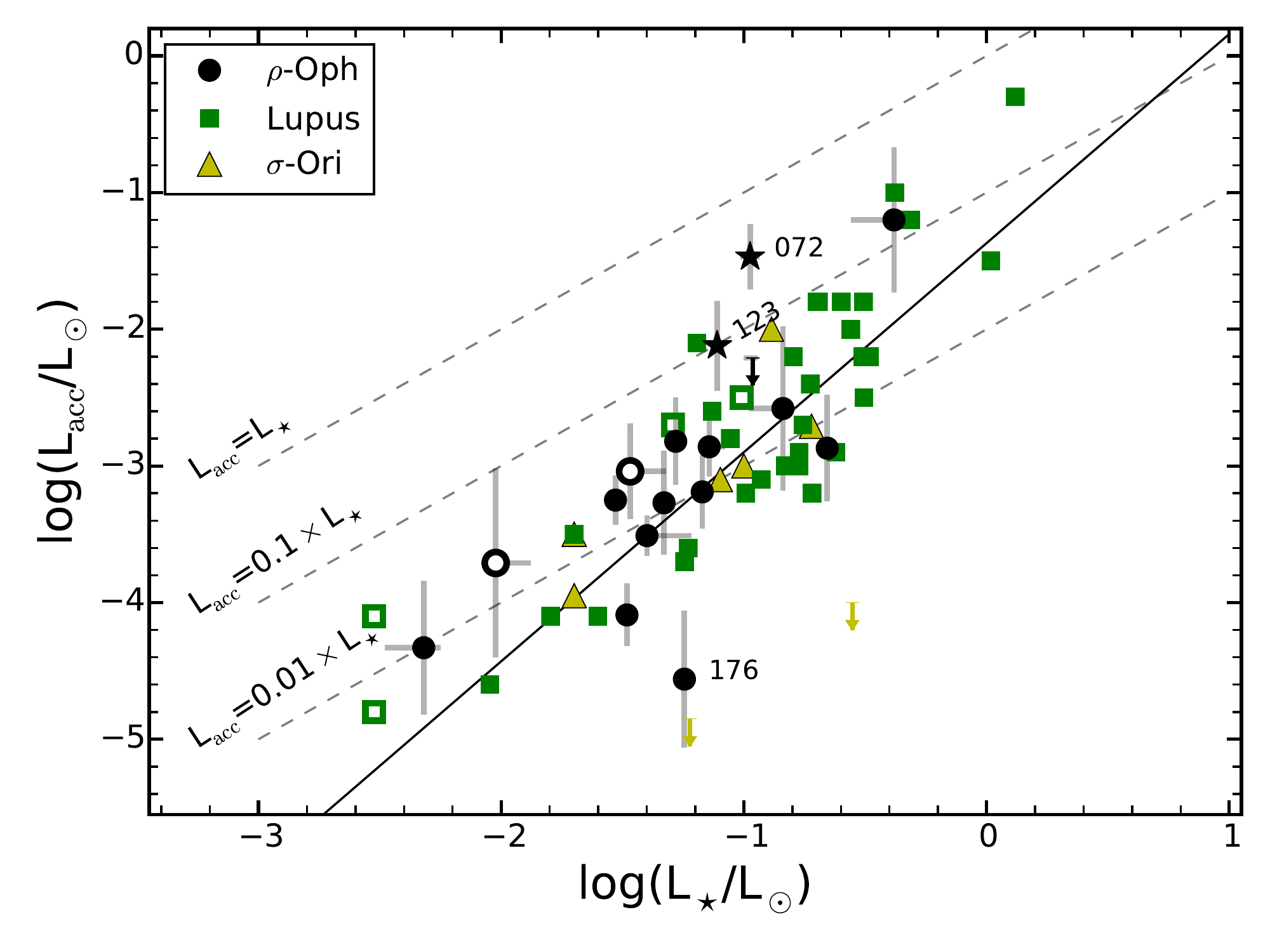}
\caption{Accretion luminosity vs stellar luminosity for the whole $\rho$-Oph sample discussed here (black symbols), for the Lupus sample of \citet{Alcala14}, reported with green symbols, and for a sample in $\sigma$-Ori studied by \citet{Rigliaco12} and reported with yellow triangles. Upper limits are shown as downward arrows. Empty symbols are used for subluminous objects, and stars for strongly veiled objects. 
Dashed lines are for different \lacc/\lstar \ ratios, going downward from 1, to 0.1, to 0.01, as labeled. The solid lines is the best fit for the GTO sample by \citet{Natta14}.
     \label{fig::lacc_lstar}}
\end{figure}

\subsection{Accretion luminosity and stellar luminosity relation}

We show in Fig.~\ref{fig::lacc_lstar} the position of our targets and those from the GTO sample (cf. legend in the plot and caption for symbols explanations) in the \lacc-\lstar \ plane. This is a particularly interesting relation as it is obtained independently from models of PMS evolution, thus it can be considered a  ``purely observational'' plot. The plot  also indicates the best-fit relation for the GTO sample by \citet{Natta14}, in which the slope is 1.53$\pm$0.18. 

All the targets analyzed here are located in the same part of the \lacc-\lstar \ plane as those from the GTO sample. In particular, we do not have any extremely strong accretors (\lacc/\lstar$>$0.5), with the only possible exception being \iso123, which is probably more intensely accreting than what is reported here. The number of objects between the \lacc = 0.01 \lstar \ and \lacc = 0.1 \lstar \ lines decreases drastically at \lstar$\lesssim$0.1 \lsun \ in the GTO sample, and the same appears to be true in our sample, as well. Also for this reason the best-fit relation has a slope significantly steeper than unity, implying that for higher luminosity targets the emission due to accretion is relatively more important than in lower \lstar \ objects. This is also confirmed  by the data collected here. 

The location of the two subluminous targets in our sample is similar to that of subluminous targets in the GTO sample (empty symbols in the plot). Similar to what was found by \citet{Alcala14} for the four subluminous targets reported here, we find that they appear in the upper boundary of the points in the \lacc-\lstar \ plane. 

Finally, our sample does not widen the spread on the \lacc-\lstar \ plane with respect to that found with the GTO sample. This suggests that the small spread in this relation found by \citet{Alcala14} is not particular to the Lupus clouds, but is probably also common  in other star-forming regions and is obtained when the data are analyzed with homogeneous methods.

\begin{figure}[!t]
\centering
\includegraphics[width=0.45\textwidth]{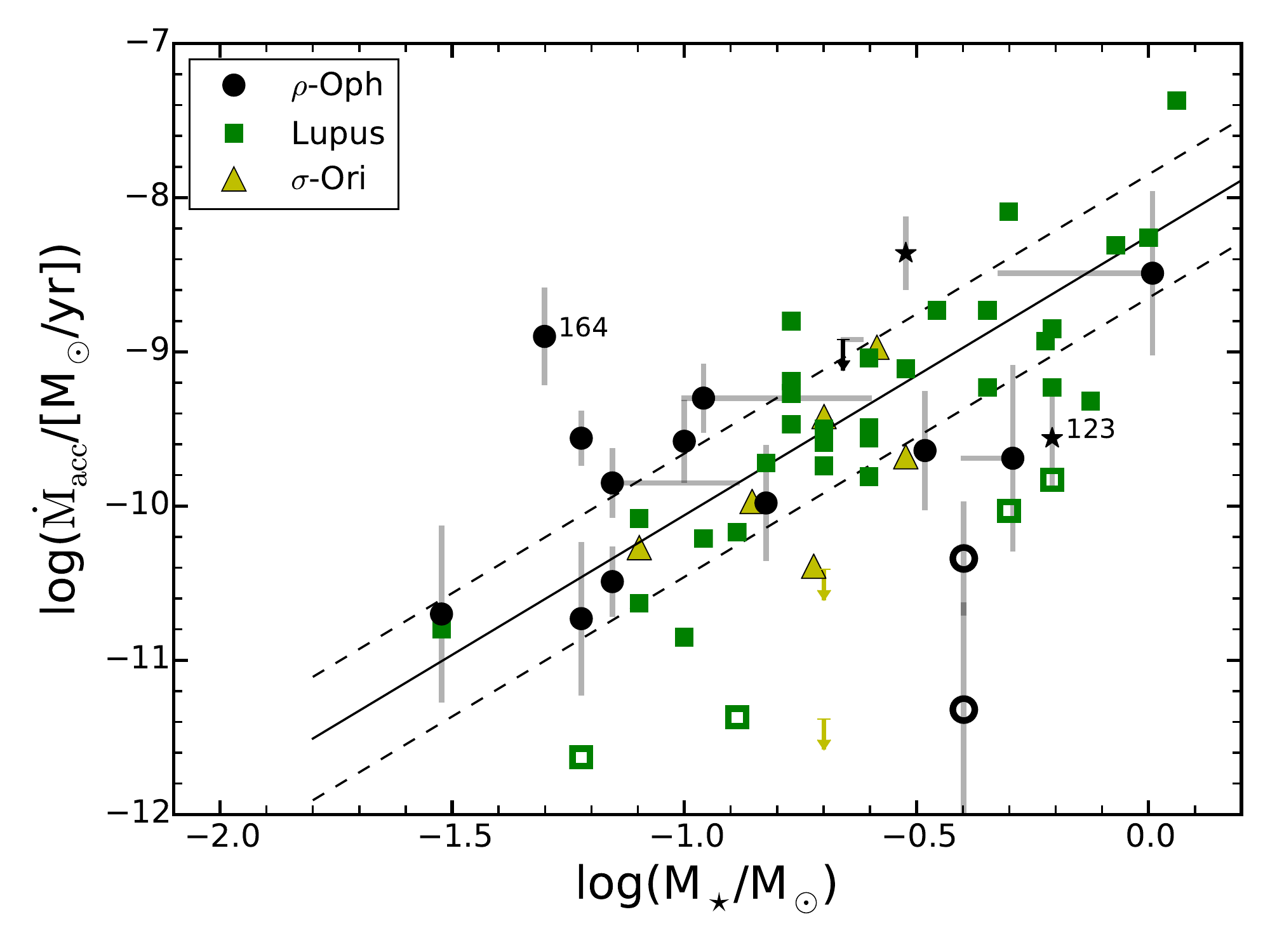}
\caption{Mass accretion rate as a function of mass for the whole $\rho$-Oph sample discussed here (black symbols) and for the Lupus sample of \citet{Alcala14}, reported with green symbols. Upper limits are shown as downward arrows. Symbols are as in Fig.~\ref{fig::lacc_lstar}. The continuous line represents the linear fit of this relation for the Lupus sample \citep{Alcala14}, and the dashed lines represent the 1\,$\sigma$ deviation from the fit.
     \label{fig::Macc_Mstar}}
\end{figure}

\subsection{Accretion as a function of stellar mass}\label{reg::sect::altogether::macc_mstar}

We present  the values of $\log$\macc \ vs $\log$\macc \ for our and the GTO samples (cf. caption and legend for explanation of the symbols) in Fig.~\ref{fig::Macc_Mstar}, together with the best-fit relation derived by \citet{Alcala14} for the Lupus sample alone. The latter has a slope of 1.8$\pm$0.2 and is particularly interesting as it is based on a large sample, but has a very small scatter ($\sigma$=0.4 dex) from the best-fit relation compared to previous studies \citep[e.g.,][]{Muzerolle03,Mohanty05,Natta06,Herczeg08,Rigliaco11a,Manara12}. Our sample does not substantially broaden the scatter of \macc \ at any \mstar, whereas most targets appear to be consistent with the best fit of the \macc-\mstar \ relation and its dispersion. Only a couple of BDs and the two objects with \mstar \ close to the hydrogen-burning limit in our sample seem to have a value of \macc \ at the upper edge of the dispersion from the best-fit relation reported by \citet{Alcala14}. However, the only one well above this relation is \iso164, which, as discussed earlier, has a slightly higher \lstar \ than objects with similar SpT. This suggests that BDs in $\rho$-Oph are not significantly stronger accretors than those in other star-forming regions, if they are strong accretors at all. As the majority of our targets follow the best-fit relation of \citet{Alcala14}, we then suggest that the scaling of \macc \ with \mstar \ is the same for VLMS and BDs. This relation is also independent of environmental conditions, as it is confirmed by our data based on targets located in a region with higher density than in the Lupus clouds.

As in the Lupus sample, the objects classified here as subluminous (empty symbols in the plot) are located in the \macc-\mstar \ plane below the others. However, these objects are probably edge-on disks, thus they do not imply that the spread is larger as their stellar and accretion parameters cannot be constrained properly with our spectra. 

Assuming that the slope of the $\log$\macc-$\log$\mstar \ relation is $\sim$1.8, we find that this is slightly shallower than that predicted by \citet{Padoan05} from models of Bondy-Hoyle accretion onto the forming YSO. This slope is instead the same as predicted by \citet{Dullemond06} assuming specific initial conditions, i.e., a small spread of angular momentum of cores at the first stages of star formation. The small spread of \macc \ at any \mstar \ found by \citet{Alcala14}, and also present in this sample, would also suggest that the initial conditions at formation are similar for these objects. As one of the stronger predictions from the latter work is a very tight \macc-$M_{\rm disk}$ relation, we await future ALMA studies coupled with large samples of objects with reliably determined stellar and accretion properties for a further confirmation of their theory. On the other hand, various studies suggested that the \macc-\mstar \ relation is mostly a result of the evolution of disks under the effects of photoevaporation \citep[e.g.,][]{Clarke&Pringle06, Ercolano14}. If this is the case, then our results seem to favor models of X-ray driven photoevaporation, which predict a slope of $\sim$1.5-1.7 \citep{Ercolano14}, with respect to those of EUV driven photoevaporation, predicting much shallower slopes of $\sim$1.35 \citep{Clarke&Pringle06}. As discussed also by \citet{Alcala14} for the Lupus sample alone, there is in the data analyzed here no evidence of the double power-law behavior suggested by \citet{Vorobyov09}, and the apparent bimodality of past data could be ascribed to mixing \macc \ determined with different methods and evolutionary models. Finally, the upper envelope of our measurements has a slope similar to the bulk of the population and steeper than 1, contrary to the prediction of \citet{Hartmann06}, which describes the \macc-\mstar \ relation as a consequence of layered disk accretion. To further constrain models using this relation, it is necessary to apply methods similar to those used for the GTO sample or to the method described here when the UV-excess is not detectable, to analyze larger and complete samples of data to avoid poor statistics. A strong caveat to this discussion is, however, that the exact value of the slope is uncertain and depends on several parameters. In particular, the choice of the evolutionary model used to derive \mstar \ is crucial, as this can lead to different slopes, and even to different spread of \macc. Here we have elected to use one evolutionary model \citep{Baraffe98}  to have results compatible with past analyses, but our results on the \macc-\mstar \ relation can be model-dependent.

Finally, we do not see in our objects, in comparison with the GTO sample, any clear evidence of dependence of \macc \ with the age of the regions. As just discussed, the objects analyzed here and those in the GTO are roughly located in the same position on the \macc-\mstar \ plane. This, however, can be the result of various effects. First of all, according to the similarity solution models by \citet{Hartmann98}, one should expect $\sim$0.5 dex difference in the values of \macc \ for objects with similar \mstar \ and with age of 1 Myr or 3 Myr. This value is comparable with the uncertainties on our \macc \ estimates, and with the spread of values observed in a given star-forming region. Then, it is also possible that an age spread of few Myr is present in the ages of objects located in the same star-forming region. This would flatten any difference in \macc \ that might, in fact, be there. Therefore,  we can only aim to constrain the dependence of \macc \ with the age of the targets with significantly larger samples and more reliable estimates for the PMS
stellar age.

\section{Conclusions}\label{sect::conclusions}

We  presented new observations of 17 VLMS and BDs with disks located in the $\rho$-Oph cluster. The spectra analyzed here were obtained with the VLT/X-Shooter instruments and were usually of good quality from $\sim$750 nm to $\sim$2500 nm, allowing us to determine SpT and $A_V$ for our targets by comparing the spectra in various molecular features with a set of photospheric templates. With this analysis, we  determined the (sub)stellar parameters for all the targets. The total sample comprises six BDs, two objects with \mstar \ close to the hydrogen burning limit, eight VLMS and low-mass YSOs, and one solar-mass PMS star. 

We determined the accretion properties for the whole sample using the intensity of the emission lines present in our spectra and compared our findings with the most complete work available in this field \citepalias{Natta06}. Our results differ from those by \citetalias{Natta06} as we find a smaller number of strong accretors, in particular, in the BD regime. This suggests that BDs in this region are not accreting at a higher rate than in other star-forming regions, contrary to previous suggestions. 

A further constraint on this result comes from the comparison of our results with those in the Lupus clouds by \citet{Alcala14} and in the $\sigma$-Ori cluster by \citet{Rigliaco12}, the GTO sample. Our objects follow the same \lacc-\lstar \ and \macc-\mstar \ relations as those in the GTO sample. However, these findings are based on very few objects.

Finally, we do not find any evidence of differences of the accretion properties of targets located in the young ($\sim$1 Myr) $\rho$-Oph cluster or in the older ($\sim$3 Myr) regions targeted in the GTO survey. However, the differences expected are small and have various effects, such as observational uncertainties and age spread, and can make this analysis even more complicated. On the other hand, this work and other recent works that have used the capabilities of the VLT/X-Shooter spectrograph together with detailed analysis techniques \citep[e.g.,][]{Rigliaco12,Alcala14,Manara14} have shown that the observational uncertainties in the estimates of accretion rates and the spread of values can be significantly reduced with respect to the past. This suggests that it will be possible with complete samples in various star-forming regions to have a clear understanding of the evolution of accretion with time and with the properties of the central PMS star.

\begin{acknowledgements}
     We thank the anonymous referee for her/his careful reading of the manuscript. We thank the ESO staff in Paranal for performing the observations in service mode. 
\end{acknowledgements}

\appendix

\section{Comments on individual objects}\label{app::ind_targ}

Here we report the previous results available in the literature in Table~\ref{tab::lit}. These are discussed in Sect.~\ref{sect::comp_lit_pars} in comparison with our estimates. In the following, we discuss some peculiar objects in the sample.

\begin{table*}
\begin{center}
\footnotesize
\caption{\label{tab::lit} Data available in the literature for the targets included in this work }
\begin{tabular}{l|cc|ccccc|c}
\hline \hline
 Object/other name &  RA(2000)  & DEC(2000) &  SpT & A$_V$ &  References & \citet{McClure10} \\ 
      &  h \, :m \, :s & $^\circ$ \, ' \, ''   &  \hbox{} & [mag] & \hbox{} & classification \\         
\hline

ISO$-$Oph023  /  SKS1 / CRBR 2317.3-1925  &  16:26:18.821  &  $-$24:26:10.52    & M7.5,M5.5,M7 & 10.0 & 1,2,3 & Disk \\
ISO$-$Oph030  /  GY5   &  16:26:21.528  &  $-$24:26:00.96    & M5.5,M6,M6 & 2.8  & 4,1,3  & Disk \\
ISO$-$Oph032  /  GY3   &  16:26:21.899  &  $-$24:44:39.76    &  M8,M7.5 & 0.0 &  4,3  & ...\\
ISO$-$Oph033  /  GY11   &  16:26:22.269  &  $-$24:24:07.06    & M8.5,M6.5,M8.5 & 8.0 &  5,1,3 & ...  \\ 
ISO$-$Oph037  /  LFAM3  /  GY21   &  16:26:23.580  &  $-$24:24:39.50    & M0,K5 & 16.1  & 7,6  & Disk/FS \\ 
ISO$-$Oph072  /  WL18    &  16:26:48.980  &  $-$24:38:25.24    & K6.5 & 10.4 &  4 & Disk \\
ISO$-$Oph087     &  16:26:58.639  &  $-$24:18:34.66    & ... & ... & ...   & ...\\
ISO$-$Oph094     &  16:27:03.591  &  $-$24:20:05.45    & M3? & 20.1 & 8   & ...\\
ISO$-$Oph102  /  GY204    &  16:27:06.596  &  $-$24:41:48.84  &  M5.5,M6 & 0.5 &  4,3  & Disk \\
ISO$-$Oph115  /  WL11  /  GY229   &  16:27:12.131  &  $-$24:34:49.14    & M0,M0 & 11.5 & 7,6  & Disk\\ 
ISO$-$Oph117  /  WLY2-32b  /  GY235   &  16:27:13.823  &  $-$24:43:31.66   & M5,K8 & 7.5 &  7,6  & Disk/FS\\ 
ISO$-$Oph123     &  16:27:17.590  &  $-$24:05:13.70    & M3.5 & 4.4 & 4 & ...\\
ISO$-$Oph160  /  B162737-241756    &  16:27:37.422  &  $-$24:17:54.87    & M6 & 6.0  & 3  & ...\\
ISO$-$Oph164  /  GY310   &  16:27:38.631  &  $-$24:38:39.19    & M8.5,M7,M6 & 5.7  &1,2,3 & Disk\\
ISO$-$Oph165  /  GY312   &  16:27:38.945  &  $-$24:40:20.67    & M2 & 15.0 & 7  & Envelope/I\\
ISO$-$Oph176  /  GY350   &  16:27:46.291  &  $-$24:31:41.19    & M6,M6 & 7.0  & 3,7 & Disk \\
ISO$-$Oph193  /  B162812-241138   &  16:28:12.720  &  $-$24:11:35.60    & M6 & 7.5 &  3 & ... \\ 

\hline

\end{tabular}
\tablefoot{ Spectral types, extinction, and mass accretion rates determined from the following studies: 1. \citet{Wilking99}; 2. \citet{Luhman99}; 3. \citet{Natta02}; 4. \citet{Wilking05}; 5. \citet{Comeron10}; 6. \citet{Gatti06}, 7. \citet[][and references therein]{McClure10}, 8.~\citet{AlvesdeOliveira12} }

\end{center}
\end{table*}

\subsection{ISO$-$Oph094}
The location of this object on the HRD that we derive is that of a very old PMS object, much older than 100 Myr. However, the spectrum of this object has SNR$\sim$0 in the whole VIS arm and also has low SNR in the NIR arm. The SpT we have derived is compatible with the highly uncertain estimates by \citet{AlvesdeOliveira12}, who suggest that this object has probably a SpT earlier than M3 and strongly extincted ($A_V$=20.1 mag). Also, their near-infrared spectra had very low SNR and their estimates are tentative. Even if we allow for higher values of $A_V$ in our fit, we obtain the same best fit as that we report in this manuscript. As there is no other information available in the literature for these targets from mid-infrared data other than \citet{Bontemps01}, we are not able to further constrain its evolutionary class, which was reported to be Class~II using ISO photometry.  With the spectra available and the information in our hand, we cannot further constrain the properties of this object. Our classification for this object as subluminous is also related to the tentative detection of forbidden emission lines in its spectrum. However, given the very low SNR of the continuum of this object, the subtraction of sky lines is not trivial and residuals are present even after careful extraction of the spectrum.

\subsection{ISO$-$Oph102}
The value of $\chi^2_{\rm like,red}$ for the best fit for this object is $\sim$12.4. As discussed in Sect.~\ref{sect::obs}, the spectrum of this target is of very poor quality in the last $\sim$100 nm of the VIS arm. We have therefore neglected some points in this region to avoid biases due to the poor quality of the spectrum. Even if the value of $\chi^2_{\rm like,red}$ is high, the best-fit template reproduces correctly the spectrum in the other features at shorter wavelength and in the NIR. Finally, our estimates are in good agreement with the literature.

\subsection{ISO$-$Oph123}
\begin{figure}[!h]
\centering
\includegraphics[height=0.7\textheight]{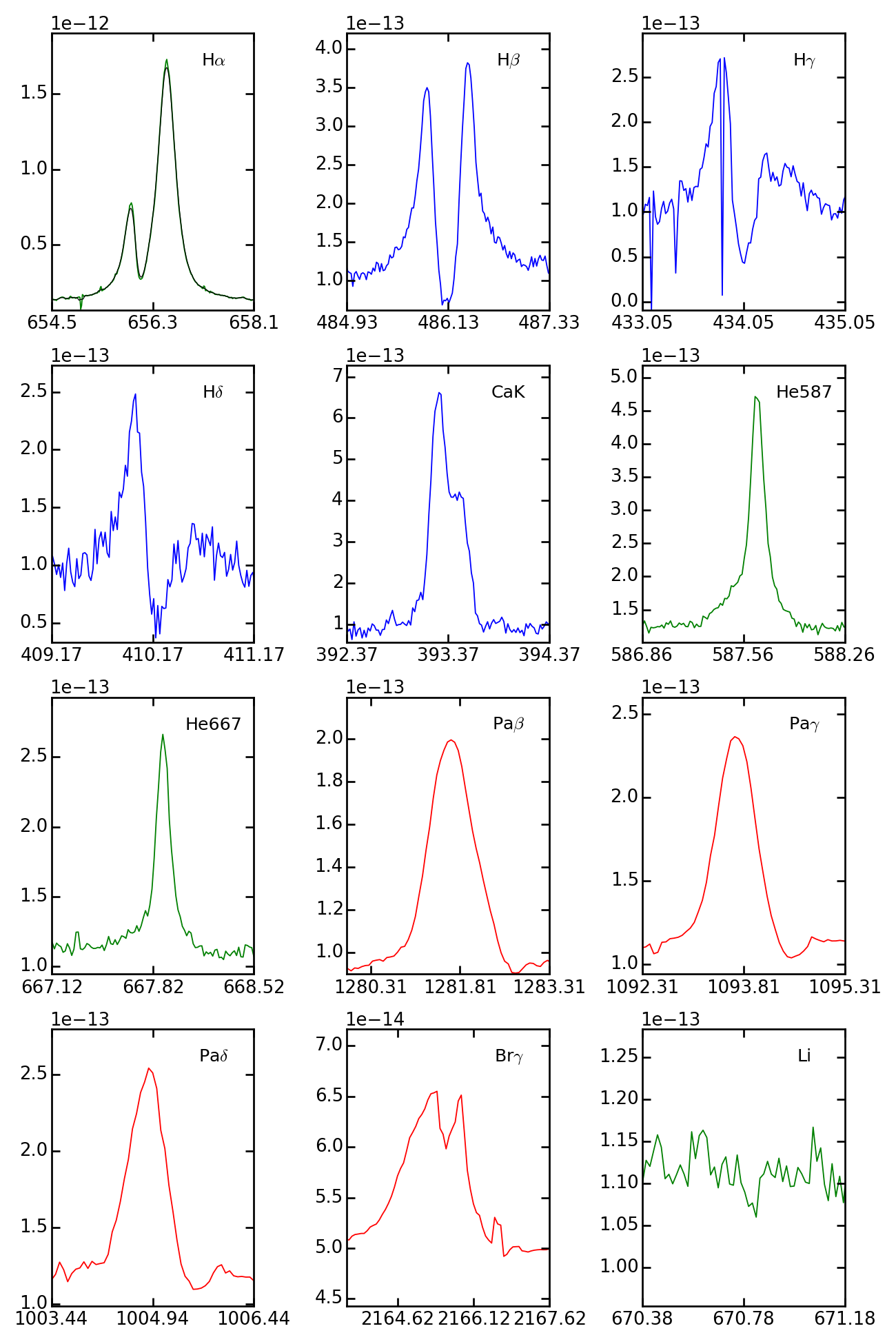}
\caption{Line profiles of permitted transition lines of ISO-Oph123. 
     \label{reg::fig::profiles1}}
\end{figure}

As  we  mentioned many times, this object is very peculiar and the analysis of its spectrum does not lead to a satisfying estimate of its stellar properties. This object has been studied by \citet{Wilking05} who reported a SpT of M3.5 with no particular comments, only that there were strong emission lines in the spectrum. \citet{Scholz12} has also noted that this object is extremely variable in the near infrared, with variations up to $\sim$0.9 mag in $K$-band. The flux calibration of the spectrum analyzed here is compatible with the 2MASS photometry in $J$- and $H$-bands, and it was only correct  by a factor 1.12 to account for slit losses  to match these bands. The magnitudes at optical wavelengths reported in the literature are, instead, slightly lower than the observed flux of this spectrum. The procedure adopted here for the rest of the targets does not result in a low value of $\chi^2_{\rm like,red}$ for this target. The best-fit estimate is also probably uncertain as this object seems a very strong accretor, and we discussed that our procedure fails if the target has \lacc/\lstar$\gtrsim$0.5, as could be the case for this object. As the excess due to accretion could significantly change  the shape of the continuum, it is not possible to determine the real reddening of the spectrum without considering all the components, such as excess due to accretion or reddening. We  tried to fit this spectrum using the procedure by \citet{Manara13b}, as the SNR in the UVB arm, although very low, is not zero. However, this automatic procedure is not able to reproduce the spectrum in the Balmer continuum region without further constraints on some parameters of the fit. We have then constrained the value of $A_V$ to be between $\sim$5 mag and $\sim$7 mag, as this is the range of values for which the spread of \lacc \ derived from the various emission lines present in the spectrum is minimum. With this constraint and forcing \lstar \ for this object to be that of a 1.2 or 2 Myr old YSO, we obtain a best fit with SpT M5. The accretion rates that would result from the fit of the UV-excess with these parameters are one order of magnitude higher than those derived from the emission lines, however, and would result in a VLMS (\mstar = 0.17 \msun) accreting at \macc$\sim$8$\times$10$^{-7}$ \msun/yr. As we show in Fig.~\ref{reg::fig::profiles1}, many emission lines in this spectrum show strong absorption features in their profiles, and this could suggest that their real flux is larger. We do not expect the difference with the real flux to be larger than a factor $\sim$2-3, which still is not compatible with the estimates from the UV-excess fit. The profiles of these lines suggest that this object might have a powerful jet, but more analyses are needed and are out of the scope of this work.

\subsection{ISO$-$Oph165}
As reported by \citet{McClure10}, this target is a Class~I object. Therefore, its photospheric continuum is strongly extincted by the envelope that still surrounds the central star and the estimates of \lstar \ are biased by this. This is the reason why this object appears to be subluminous on the HRD.

\clearpage

\Online

\section{Best fit of observed $\rho$-Oph spectra with templates}\label{app::plot_spectra}

\begin{landscape}

\begin{figure}[]
\centering
\includegraphics[height=0.93\textwidth]{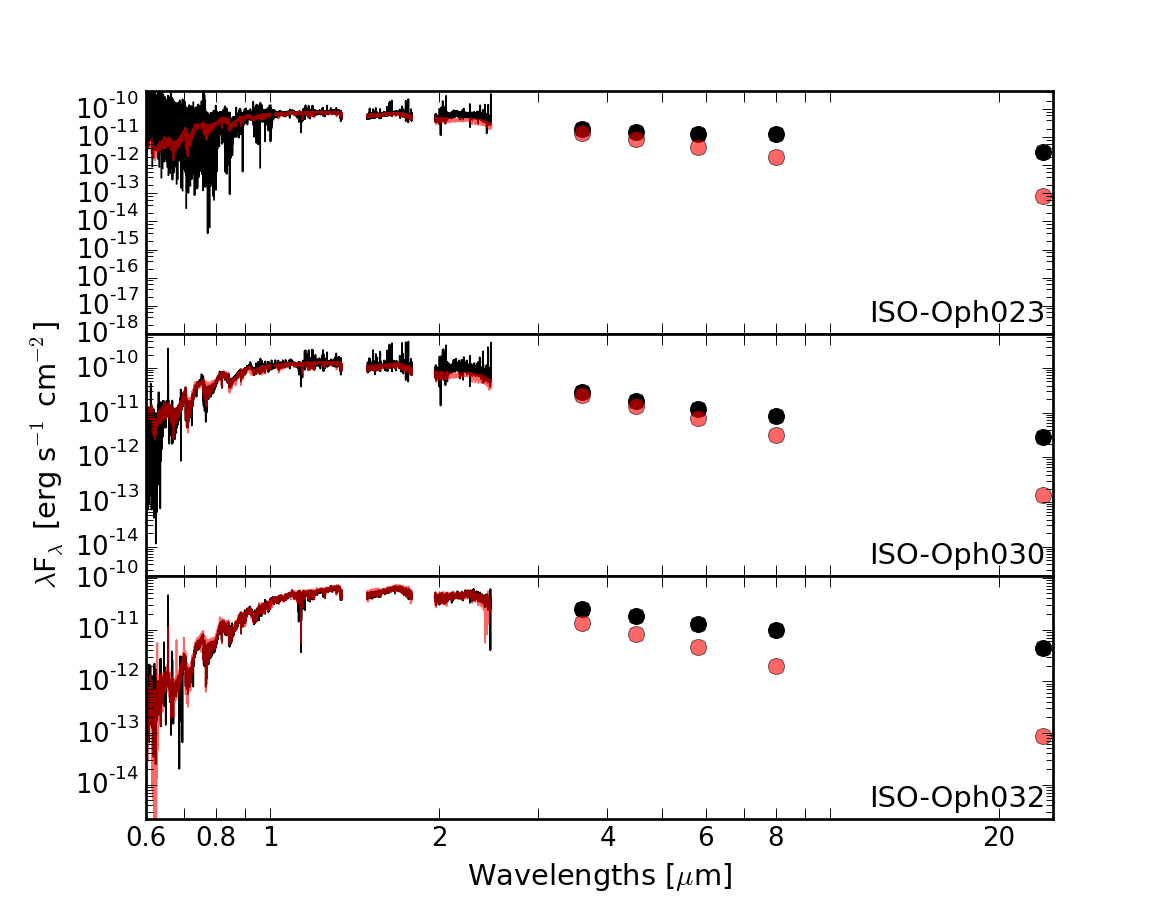}
\caption{Spectra of $\rho$-Oph targets from 600 to 2450 nm and \textit{Spitzer} photometry up to 24$\mu$m. The extinction-corrected spectra and photometric points for the targets are shown in black. Red lines and points are used for the best-fit template spectrum, which is normalized at $\lambda$=1025 nm to match the extinction-corrected target spectrum. Veiling due to accretion or disk emission is not included.
     \label{reg::fig::best_fit_oph}}
\end{figure}
\end{landscape}

\begin{landscape}
\begin{figure}[]
\centering
\includegraphics[height=0.93\textwidth]{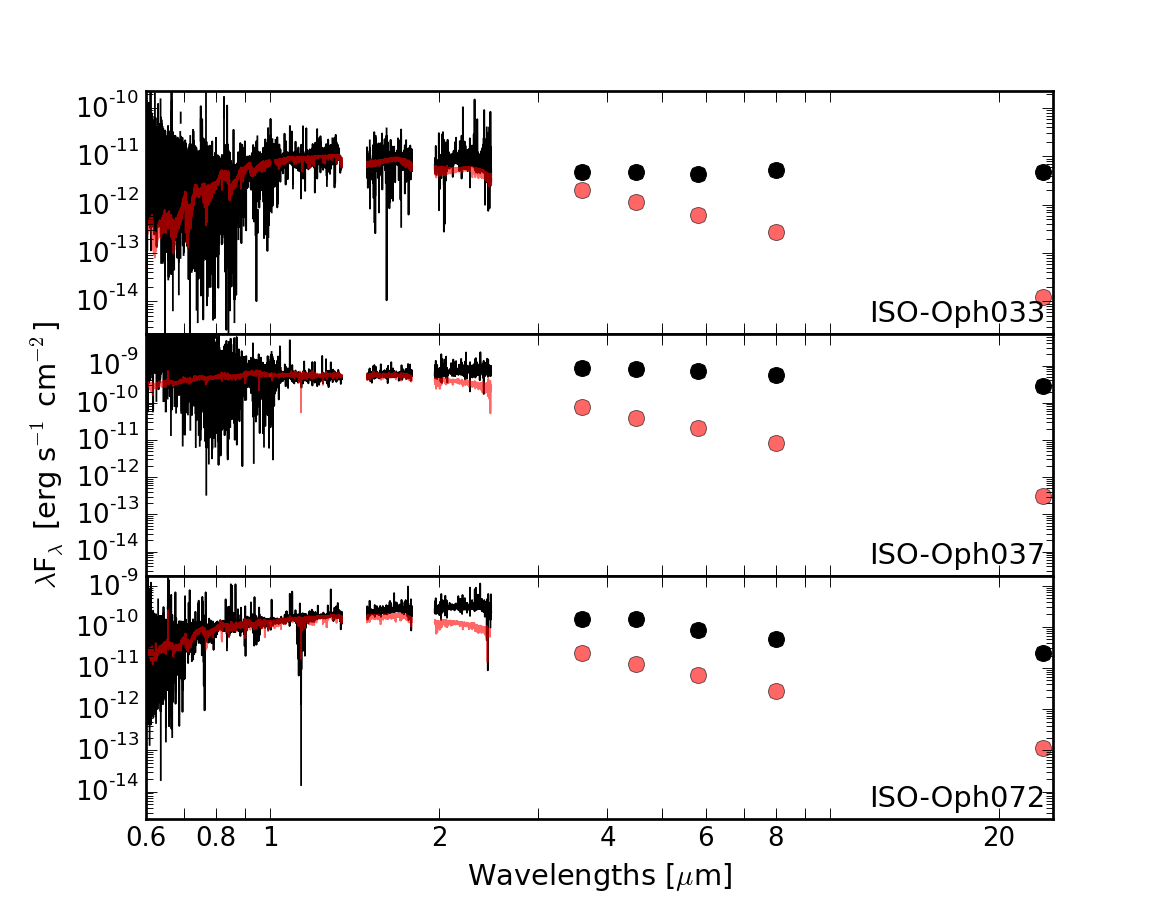}
\caption{Same as Fig.~\ref{reg::fig::best_fit_oph}. 
     \label{reg::fig::best_fit_oph2}}
\end{figure}
\end{landscape}

\begin{landscape}
\begin{figure}[]
\centering
\includegraphics[height=0.93\textwidth]{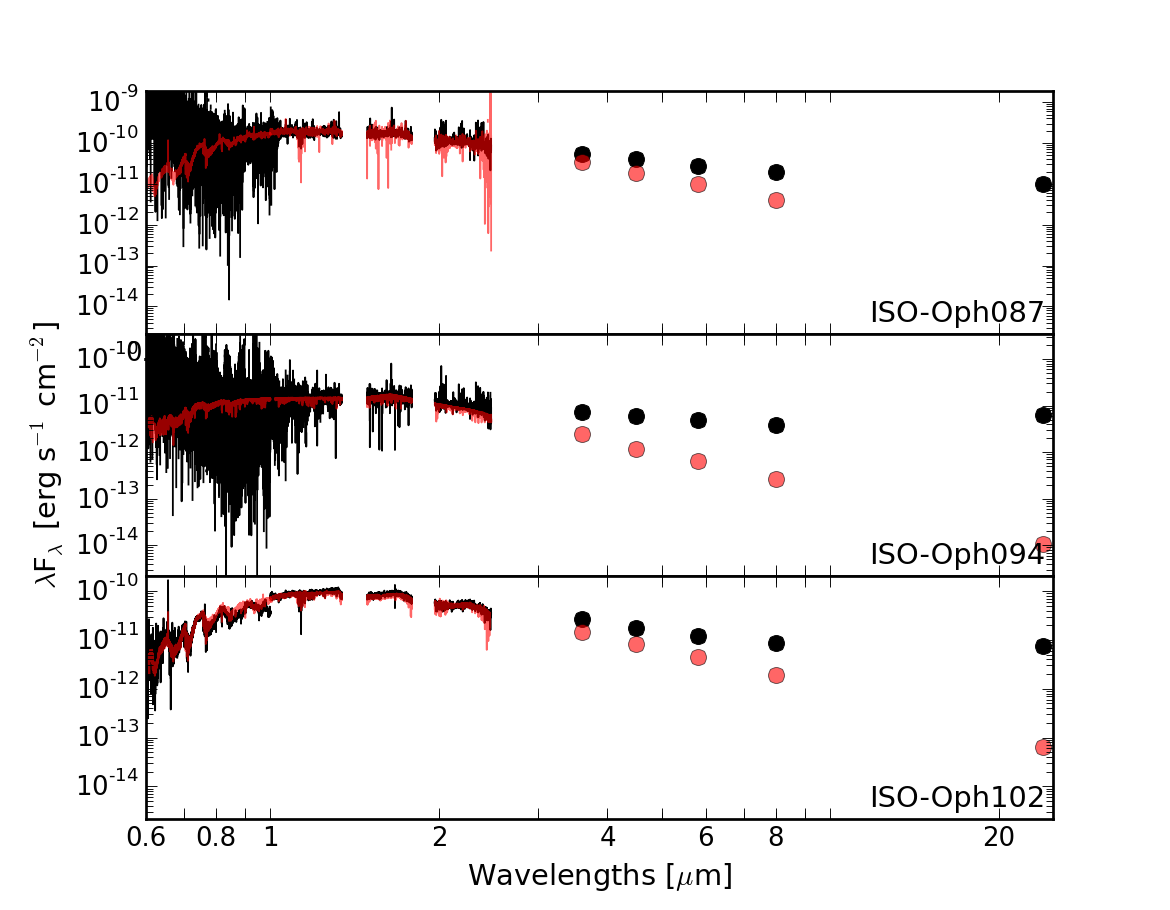}
\caption{Same as Fig.~\ref{reg::fig::best_fit_oph}. 
     \label{reg::fig::best_fit_oph3}}
\end{figure}
\end{landscape}

\begin{landscape}
\begin{figure}[]
\centering
\includegraphics[height=0.93\textwidth]{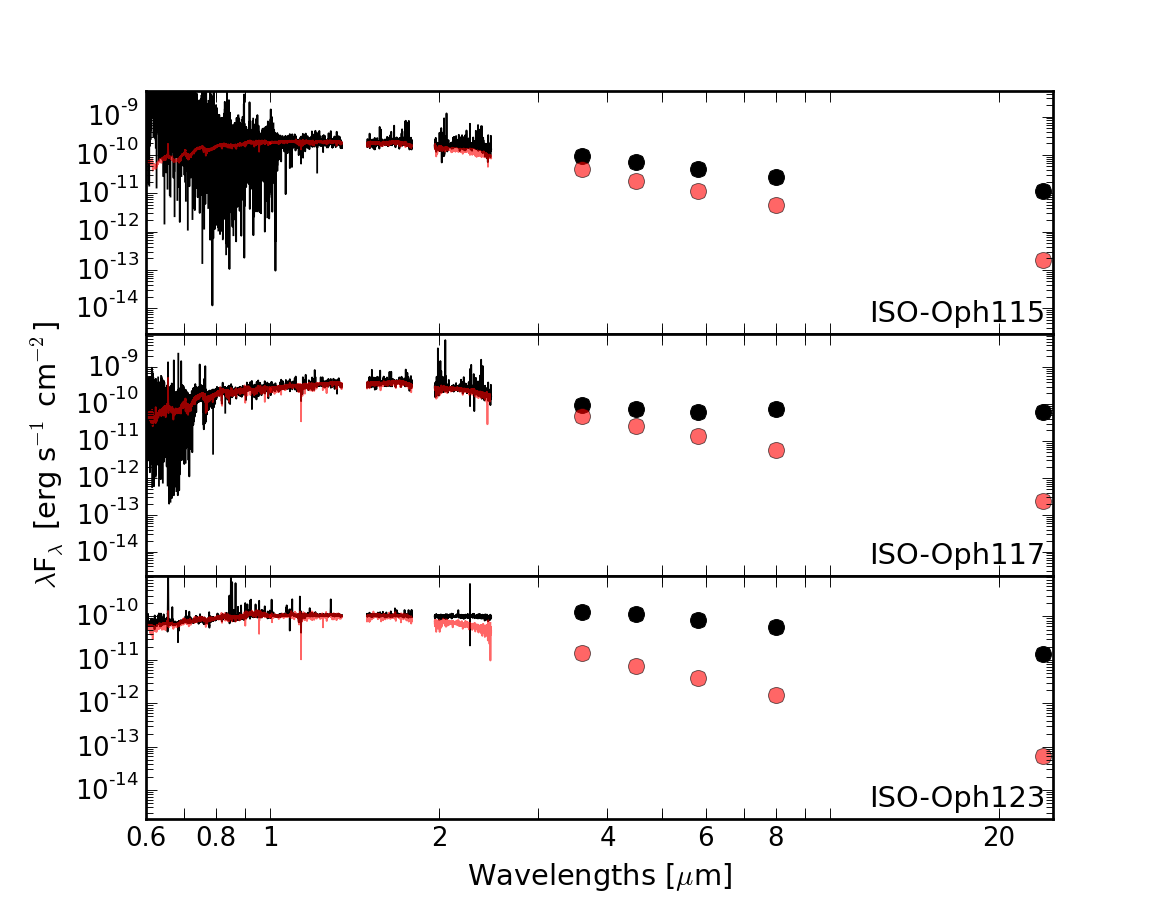}
\caption{Same as Fig.~\ref{reg::fig::best_fit_oph}. 
     \label{reg::fig::best_fit_oph4}}
\end{figure}
\end{landscape}

\begin{landscape}
\begin{figure}[]
\centering
\includegraphics[height=0.93\textwidth]{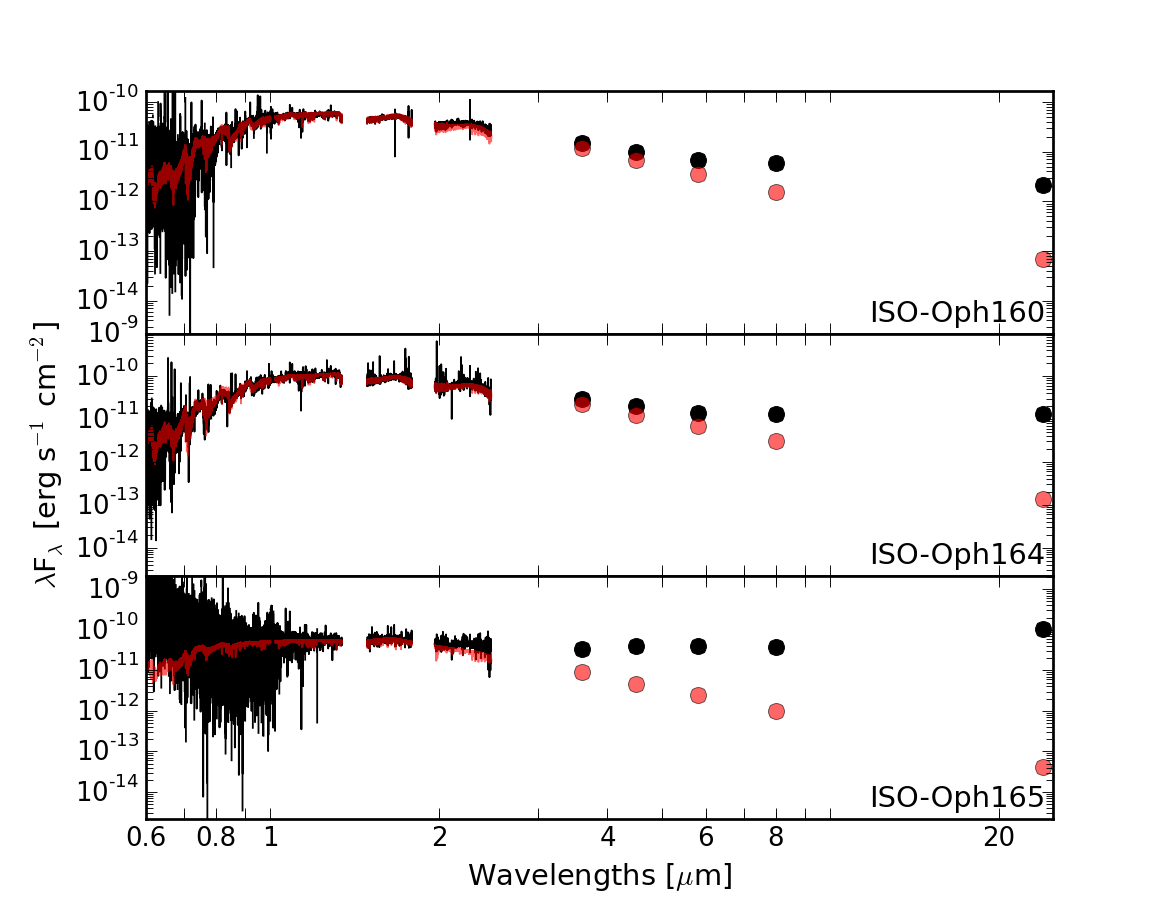}
\caption{Same as Fig.~\ref{reg::fig::best_fit_oph}. 
     \label{reg::fig::best_fit_oph5}}
\end{figure}

\end{landscape}

\begin{landscape}
\begin{figure}[]
\centering
\includegraphics[height=0.93\textwidth]{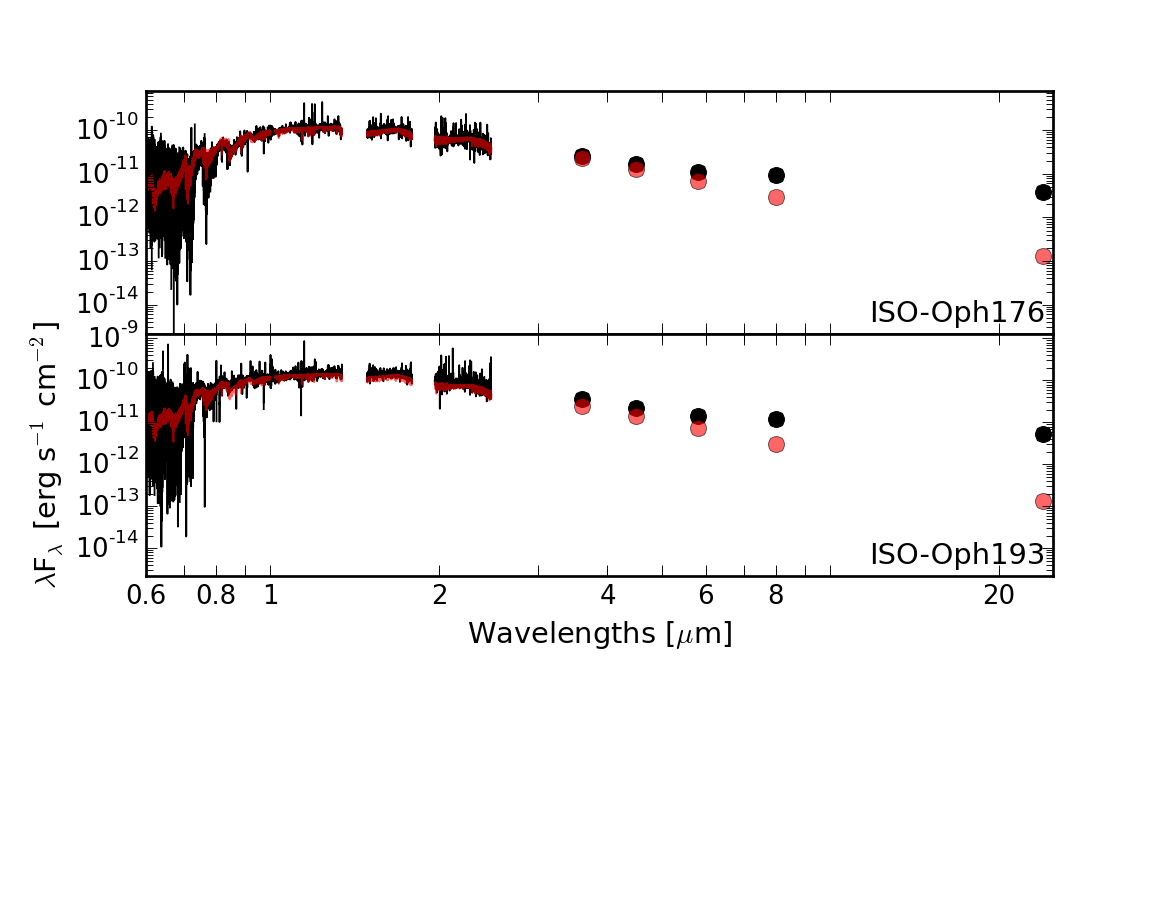}
\caption{Same as Fig.~\ref{reg::fig::best_fit_oph}. 
     \label{reg::fig::best_fit_oph6}}
\end{figure}

\end{landscape}

\section{Permitted emission line profiles in the X-Shooter spectra}\label{app::plot_spectra_lines}

\begin{landscape} 
\begin{table} 
\centering 
\footnotesize 
\caption{\label{tab::perm_line_flux}Extinction-corrected fluxes and equivalent widths of Balmer, Paschen, and Bracket lines} 
\begin{tabular}{l|r|r|r|r|r|r|r|r} 
\hline\hline 
Object & $f_{\rm H\alpha}$ & $EW_{\rm H\alpha}$ & $f_{\rm H\beta}$ & $EW_{\rm H\beta}$  & $f_{\rm H\gamma}$ & $EW_{\rm H\gamma}$  & $f_{\rm H\delta}$ & $EW_{\rm H\delta}$  \\ 
\hbox{} &  \hbox{} & [nm]&  \hbox{} & [nm]&  \hbox{} & [nm]&  \hbox{} & [nm] \\ 
\hline 
ISO$-$Oph023 & 10.3 $\pm$ 4.2 & -9.472 $\pm$ 3.864 & $<$105.0 & ... & $<$175.0 & ... & $<$231.0 & ... \\ 
ISO$-$Oph030 & 18.2 $\pm$ 0.3 & -8.523 $\pm$ 0.162 & 1.2 $\pm$ 0.3 & -3.303 $\pm$ 0.972 & $<$0.7 & ... & $<$0.9 & ... \\ 
ISO$-$Oph032 & 2.460 $\pm$ 0.030 & -13.535 $\pm$ 0.165 & 0.269 $\pm$ 0.018 & -7.954 $\pm$ 0.542 & 0.141 $\pm$ 0.016 & -8.077 $\pm$ 0.901 & 0.105 $\pm$ 0.016 & -6.615 $\pm$ 1.003 \\ 
ISO$-$Oph033 & 1.3 $\pm$ 0.4 & -12.118 $\pm$ 3.614 & $<$20.0 & ... & $<$26.9 & ... & $<$38.2 & ... \\ 
ISO$-$Oph037 & 3430.0 $\pm$ 795.0 & -4.534 $\pm$ 1.052 & ... & ... & ... & ... & ... & ... \\ 
ISO$-$Oph072 & 234.0 $\pm$ 6.2 & -14.248 $\pm$ 0.379 & 45.6 $\pm$ 14.0 & -2.297 $\pm$ 0.704 & $<$45.2 & ... & $<$72.3 & ... \\ 
ISO$-$Oph087 & $<$101.0 & ... & $<$2490.0 & ... & $<$4540.0 & ... & $<$7760.0 & ... \\ 
ISO$-$Oph094 & $<$9.2 & ... & $<$116.0 & ... & $<$168.0 & ... & $<$226.0 & ... \\ 
ISO$-$Oph102 & 8.8 $\pm$ 0.3 & -4.760 $\pm$ 0.146 & 0.6 $\pm$ 0.2 & -2.400 $\pm$ 0.621 & 0.8 $\pm$ 0.3 & -3.769 $\pm$ 1.273 & 0.7 $\pm$ 0.3 & -2.054 $\pm$ 0.834 \\ 
ISO$-$Oph115 & $<$408.0 & ... & $<$14200.0 & ... & $<$47600.0 & ... & $<$92700.0 & ... \\ 
ISO$-$Oph117 & 75.9 $\pm$ 8.0 & -4.468 $\pm$ 0.472 & $<$42.8 & ... & $<$184.0 & ... & $<$146.0 & ... \\ 
ISO$-$Oph123 & 81.6 $\pm$ 0.9 & -6.068 $\pm$ 0.066 & 12.0 $\pm$ 0.8 & -1.232 $\pm$ 0.080 & 4.9 $\pm$ 1.1 & -0.553 $\pm$ 0.119 & 3.0 $\pm$ 0.6 & -0.377 $\pm$ 0.075 \\ 
ISO$-$Oph160 & 8.5 $\pm$ 0.7 & -10.546 $\pm$ 0.883 & $<$3.8 & ... & $<$7.6 & ... & $<$8.9 & ... \\ 
ISO$-$Oph164 & 10.5 $\pm$ 0.4 & -6.335 $\pm$ 0.222 & 2.8 $\pm$ 0.6 & -3.133 $\pm$ 0.631 & 1.5 $\pm$ 0.6 & -1.881 $\pm$ 0.724 & $<$1.7 & ... \\ 
ISO$-$Oph165 & 36.0 $\pm$ 22.0 & -6.484 $\pm$ 3.966 & $<$1130.0 & ... & $<$2860.0 & ... & $<$5170.0 & ... \\ 
ISO$-$Oph176 & 0.8 $\pm$ 0.3 & -0.728 $\pm$ 0.232 & ... & ... & ... & ... & ... & ... \\ 
ISO$-$Oph193 & 43.7 $\pm$ 2.6 & -9.836 $\pm$ 0.592 & ... & ... & ... & ... & ... & ... \\ 
\hline 
\end{tabular} 
\begin{tabular}{l|r|r|r|r|r|r|r|r} 
\hline\hline 
Object  & $f_{\rm Pa\beta}$ & $EW_{\rm Pa\beta}$  & $f_{\rm Pa\gamma}$ & $EW_{\rm Pa\gamma}$  & $f_{\rm Pa\delta}$ & $EW_{\rm Pa\delta}$ & $f_{\rm Br\gamma}$ & $EW_{\rm Br\gamma}$ \\ 
\hbox{} &  \hbox{} & [nm]&  \hbox{} & [nm]&  \hbox{} & [nm]&  \hbox{} & [nm]\\ 
\hline 
ISO$-$Oph023 & 0.3 $\pm$ 0.2 & -0.039 $\pm$ 0.035 & 0.4 $\pm$ 0.2 & -0.060 $\pm$ 0.032 & $<$1.3 & ... & 0.2 $\pm$ 0.1 & -0.053 $\pm$ 0.026  \\ 
ISO$-$Oph030 & 0.4 $\pm$ 0.2 & -0.033 $\pm$ 0.023 & 0.6 $\pm$ 0.3 & -0.051 $\pm$ 0.022 & 1.1 $\pm$ 0.5 & -0.107 $\pm$ 0.049 & $<$0.1 & ...  \\ 
ISO$-$Oph032 & 0.069 $\pm$ 0.109 & -0.014 $\pm$ 0.022 & 0.278 $\pm$ 0.137 & -0.062 $\pm$ 0.030 & $<$0.13 & ... & $<$0.05 & ...  \\ 
ISO$-$Oph033 & $<$0.1 & ... & $<$0.1 & ... & $<$0.5 & ... & $<$0.0 & ...  \\ 
ISO$-$Oph037 & 24.5 $\pm$ 2.0 & -0.614 $\pm$ 0.049 & 23.6 $\pm$ 3.8 & -0.468 $\pm$ 0.075 & $<$38.7 & ... & $<$1.1 & ...  \\ 
ISO$-$Oph072 & 39.5 $\pm$ 0.8 & -2.455 $\pm$ 0.052 & 36.5 $\pm$ 1.1 & -2.088 $\pm$ 0.063 & 21.7 $\pm$ 2.4 & -1.485 $\pm$ 0.166 & 14.1 $\pm$ 0.3 & -0.946 $\pm$ 0.022  \\ 
ISO$-$Oph087 & $<$0.9 & ... & $<$1.0 & ... & $<$10.9 & ... & $<$0.4 & ...  \\ 
ISO$-$Oph094 & 0.2 $\pm$ 0.1 & -0.207 $\pm$ 0.122 & $<$0.3 & ... & $<$2.8 & ... & $<$0.0 & ...  \\ 
ISO$-$Oph102 & $<$0.2 & ... & 0.4 $\pm$ 0.2 & -0.044 $\pm$ 0.021 & $<$0.3 & ... & $<$0.1 & ...  \\ 
ISO$-$Oph115 & 3.0 $\pm$ 1.2 & -0.144 $\pm$ 0.058 & $<$2.0 & ... & $<$20.8 & ... & $<$0.6 & ...  \\ 
ISO$-$Oph117 & 0.9 $\pm$ 0.8 & -0.030 $\pm$ 0.028 & 0.9 $\pm$ 0.9 & -0.031 $\pm$ 0.031 & $<$2.3 & ... & $<$0.4 & ...  \\ 
ISO$-$Oph123 & 9.0 $\pm$ 0.2 & -0.980 $\pm$ 0.020 & 7.7 $\pm$ 0.4 & -0.701 $\pm$ 0.040 & 7.7 $\pm$ 0.6 & -0.660 $\pm$ 0.055 & 2.1 $\pm$ 0.1 & -0.416 $\pm$ 0.018  \\ 
ISO$-$Oph160 & 1.1 $\pm$ 0.1 & -0.238 $\pm$ 0.032 & 0.9 $\pm$ 0.2 & -0.178 $\pm$ 0.046 & $<$0.6 & ... & 0.2 $\pm$ 0.0 & -0.101 $\pm$ 0.023  \\ 
ISO$-$Oph164 & 1.5 $\pm$ 0.2 & -0.160 $\pm$ 0.018 & 4.4 $\pm$ 0.7 & -0.368 $\pm$ 0.056 & 1.9 $\pm$ 0.9 & -0.219 $\pm$ 0.101 & $<$0.2 & ...  \\ 
ISO$-$Oph165 & 0.5 $\pm$ 0.3 & -0.116 $\pm$ 0.069 & $<$0.6 & ... & $<$9.0 & ... & $<$0.1 & ...  \\ 
ISO$-$Oph176 & $<$0.3 & ... & $<$0.2 & ... & $<$0.8 & ... & $<$0.1 & ...  \\ 
ISO$-$Oph193 & 1.6 $\pm$ 0.6 & -0.138 $\pm$ 0.048 & 1.0 $\pm$ 0.5 & -0.076 $\pm$ 0.037 & $<$2.9 & ... & $<$0.1 & ...  \\ 
\hline 
\end{tabular} 
\tablefoot{Fluxes are reported in units of 10$^{-14}$ erg s$^{-1}$ cm$^{-2}$ followed by their errors. Upper limits (3 $\sigma$) are reported with $<$. }
\end{table} 
\end{landscape} 

\begin{landscape} 
\begin{table} 
\centering 
\footnotesize 
\caption{\label{tab::perm_line_flux2}Extinction-corrected fluxes and equivalent widths of helium and calcium lines} 
\begin{tabular}{l|r|r|r|r|r|r} 
\hline\hline 
Object & $f_{\rm He \lambda 587.6}$ & $EW_{\rm He \lambda 587.6}$ & $f_{\rm He \lambda 667.8}$ & $EW_{\rm He \lambda 667.8}$  & $f_{\rm CaK}$ & $EW_{\rm CaK}$ \\ 
\hbox{} &  \hbox{} & [nm]&  \hbox{} & [nm]&  \hbox{} & [nm] \\ 
\hline 
ISO$-$Oph023 & $<$28.6 & ... & $<$7.6 & ... & $<$248.0 & ...  \\ 
ISO$-$Oph030 & 0.3 $\pm$ 0.1 & -0.432 $\pm$ 0.190 & 0.2 $\pm$ 0.1 & -0.141 $\pm$ 0.050 & $<$0.7 & ...  \\ 
ISO$-$Oph032 & 0.054 $\pm$ 0.007 & -1.162 $\pm$ 0.148 & 0.029 $\pm$ 0.006 & -0.277 $\pm$ 0.060 & 0.036 $\pm$ 0.011 & -4.716 $\pm$ 1.382  \\ 
ISO$-$Oph033 & $<$8.4 & ... & $<$1.4 & ... & $<$33.2 & ...  \\ 
ISO$-$Oph037 & $<$8500.0 & ... & $<$1390.0 & ... & ... & ...  \\ 
ISO$-$Oph072 & $<$11.0 & ... & $<$3.4 & ... & $<$99.7 & ...  \\ 
ISO$-$Oph087 & $<$663.0 & ... & $<$106.0 & ... & $<$9210.0 & ...  \\ 
ISO$-$Oph094 & $<$104.0 & ... & $<$9.3 & ... & $<$240.0 & ...  \\ 
ISO$-$Oph102 & 0.6 $\pm$ 0.1 & -0.875 $\pm$ 0.190 & 0.3 $\pm$ 0.0 & -0.234 $\pm$ 0.038 & 0.7 $\pm$ 0.3 & -3.602 $\pm$ 1.470  \\ 
ISO$-$Oph115 & $<$2380.0 & ... & $<$550.0 & ... & $<$127000.0 & ...  \\ 
ISO$-$Oph117 & $<$27.1 & ... & $<$9.7 & ... & $<$192.0 & ...  \\ 
ISO$-$Oph123 & 5.4 $\pm$ 0.3 & -0.472 $\pm$ 0.025 & 2.0 $\pm$ 0.1 & -0.194 $\pm$ 0.014 & 15.6 $\pm$ 0.7 & -1.872 $\pm$ 0.085  \\ 
ISO$-$Oph160 & $<$2.4 & ... & $<$2.7 & ... & $<$24.1 & ...  \\ 
ISO$-$Oph164 & 0.7 $\pm$ 0.2 & -1.356 $\pm$ 0.462 & 0.3 $\pm$ 0.1 & -0.362 $\pm$ 0.105 & $<$2.0 & ...  \\ 
ISO$-$Oph165 & $<$569.0 & ... & $<$66.2 & ... & $<$6960.0 & ...  \\ 
ISO$-$Oph176 & $<$3.4 & ... & $<$1.4 & ... & ... & ...  \\ 
ISO$-$Oph193 & $<$6.9 & ... & $<$2.3 & ... & ... & ...  \\ 
\hline 
\end{tabular} 
\tablefoot{Fluxes are reported in units of 10$^{-14}$ erg s$^{-1}$ cm$^{-2}$ followed by their errors. Upper limits (3 $\sigma$) are reported with $<$. }
\end{table} 
\end{landscape}

\onecolumn
\section{Stellar and accretion parameters by \citet{Natta06} corrected for the most recent distance estimate}

\begin{longtable}{clccccccccc}  
\caption{\label{tab::natta_125}Properties of $\rho$-Ophiuci young stellar objects from \citetalias{Natta06} using distance of 125 pc.} \\ 
\hline\hline 
\# & Object & Class & log$L_{\star}$ & log$T_{\rm eff}$ & log$M_\star$ & Line & EW(line) & log$L_{\rm line}$ & log$L_{\rm acc}$  & log$\dot{M}_{\rm acc}$ \\ 
\hbox{} & (ISO\#) & \hbox{}  & [$L_{\odot}$] & [K] & [M$_\odot$] & \hbox{} & [\AA] & [$L_{\odot}$] & [$L_{\odot}$] & [$M_\odot$/yr]  \\ 
\hline 
\endfirsthead 
\caption{Continued.} \\ 
\hline\hline 
\# & Object & Class & log$L_{\star}$ & log$T_{\rm eff}$ & log$M_\star$ & Line & EW(line) & log$L_{\rm line}$ & log$L_{\rm acc}$  & log$\dot{M}_{\rm acc}$ \\ 
\hbox{} & (ISO\#) & \hbox{}  & [$L_{\odot}$] & [K] & [M$_\odot$] & \hbox{} & [\AA] & [$L_{\odot}$] & [$L_{\odot}$] & [$M_\odot$/yr]  \\ 
\hline 
\endhead 
\hline 
\endfoot 
1 & ISO$-$Oph001 & II & 0.16 & 3.58 & 0.02 & Pa$\beta$ & $<$0.7 & $<$-4.36 & $<$-1.93  & $<$-9.01 \\ 
2 & ISO$-$Oph002 & II & -0.21 & 3.54 & -0.24 & Pa$\beta$ & 2.8 & -4.14 & -1.63  & -8.55 \\ 
3 & ISO$-$Oph003 & II & -0.13 & 3.55 & -0.19 & Pa$\beta$ & 3.9 & -3.92 & -1.33  & -8.28 \\ 
4 & ISO$-$Oph005 & III & 0.12 & 3.58 & -0.01 & Pa$\beta$ & $<$0.5 & $<$-4.57 & $<$-2.21  & $<$-9.27 \\ 
5 & ISO$-$Oph006 & II & 0.07 & 3.58 & -0.03 & Pa$\beta$ & 19.0 & -3.03 & -0.12  & -7.18 \\ 
6 & ISO$-$Oph009 & II & -1.13 & 3.48 & -0.99 & Pa$\beta$ & $<$0.3 & $<$-5.84 & $<$-3.94  & $<$-10.46 \\ 
7 & ISO$-$Oph011 & III & -0.29 & 3.53 & -0.31 & Pa$\beta$ & $<$0.8 & $<$-4.61 & $<$-2.26  & $<$-9.14 \\ 
8 & ISO$-$Oph012 & II & -1.07 & 3.48 & -0.95 & Pa$\beta$ & $<$0.2 & $<$-6.05 & $<$-4.23  & $<$-10.75 \\ 
9 & ISO$-$Oph013 & II & -0.26 & 3.54 & -0.28 & Pa$\beta$ & 2.1 & -4.31 & -1.86  & -8.76 \\ 
10 & ISO$-$Oph014 & III & 0.01 & 3.57 & -0.09 & Pa$\beta$ & $<$0.5 & $<$-4.71 & $<$-2.40  & $<$-9.41 \\ 
11 & ISO$-$Oph016 & III & 0.49 & 3.66 & 0.20 & Pa$\beta$ & -5.7 & ... & ...  & ... \\ 
12 & ISO$-$Oph017 & II & 0.54 & 3.67 & 0.22 & Pa$\beta$ & $<$0.8 & $<$-4.08 & $<$-1.55  & $<$-8.81 \\ 
13 & ISO$-$Oph018 & III & 0.06 & 3.58 & -0.04 & Pa$\beta$ & $<$0.5 & $<$-4.67 & $<$-2.36  & $<$-9.41 \\ 
14 & ISO$-$Oph019 & II & 0.55 & 3.67 & 0.23 & Pa$\beta$ & $<$0.9 & $<$-3.97 & $<$-1.40  & $<$-8.67 \\ 
15 & ISO$-$Oph020 & II & -0.02 & 3.56 & -0.11 & Pa$\beta$ & 1.2 & -4.30 & -1.85  & -8.85 \\ 
16 & ISO$-$Oph023 & II & -1.85 & 3.45 & -1.36 & Pa$\beta$ & 1.8 & -5.63 & -3.65  & -10.09 \\ 
17 & ISO$-$Oph024 & II & 0.26 & 3.60 & 0.07 & Pa$\beta$ & 8.9 & -3.21 & -0.36  & -7.47 \\ 
18 & ISO$-$Oph026 & II & -0.48 & 3.52 & -0.48 & Pa$\beta$ & 2.4 & -4.27 & -1.80  & -8.58 \\ 
19 & ISO$-$Oph028 & III & 0.36 & 3.61 & 0.12 & Pa$\beta$ & $<$0.6 & $<$-4.33 & $<$-1.89  & $<$-9.03 \\ 
20 & ISO$-$Oph030 & II & -1.42 & 3.47 & -1.14 & Pa$\beta$ & 0.3 & -6.03 & -4.20  & -10.68 \\ 
21 & ISO$-$Oph032 & II & -1.71 & 3.45 & -1.29 & Pa$\beta$ & 0.4 & -6.13 & -4.34  & -10.79 \\ 
22 & ISO$-$Oph033 & II & -3.26 & 3.33 & nan & Pa$\beta$ & $<$0.7 & $<$-7.22 & $<$-5.82  & $<$nan \\ 
23 & ISO$-$Oph035 & II & -1.17 & 3.48 & -1.01 & Pa$\beta$ & -0.6 & ... & ...  & ... \\ 
24 & ISO$-$Oph036 & II & 0.70 & 3.68 & 0.31 & Pa$\beta$ & 0.7 & -3.92 & -1.34  & -8.64 \\ 
25 & ISO$-$Oph037 & II & -0.79 & 3.50 & -0.78 & Pa$\beta$ & 3.0 & -4.62 & -2.28  & -8.87 \\ 
26 & ISO$-$Oph038 & II & 0.00 & 3.56 & -0.10 & Pa$\beta$ & $<$0.3 & $<$-4.87 & $<$-2.63  & $<$-9.64 \\ 
27 & ISO$-$Oph039 & II & 0.99 & 3.71 & 0.37 & Pa$\beta$ & $<$0.5 & $<$-3.87 & $<$-1.26  & $<$-8.54 \\ 
28 & ISO$-$Oph040 & II & 0.53 & 3.66 & 0.22 & Pa$\beta$ & 12.7 & -2.83 & 0.15  & -7.11 \\ 
29 & ISO$-$Oph041 & II & -0.61 & 3.51 & -0.61 & Pa$\beta$ & $<$0.3 & $<$-5.42 & $<$-3.36  & $<$-10.06 \\ 
30 & ISO$-$Oph043 & II & 0.01 & 3.57 & -0.09 & Pa$\beta$ & $<$1.0 & $<$-4.40 & $<$-1.99  & $<$-9.00 \\ 
31 & ISO$-$Oph044 & III & -0.15 & 3.55 & -0.21 & Pa$\beta$ & $<$0.7 & $<$-4.69 & $<$-2.38  & $<$-9.32 \\ 
32 & ISO$-$Oph046 & II & -0.11 & 3.55 & -0.18 & Pa$\beta$ & 16.5 & -3.29 & -0.48  & -7.43 \\ 
33 & ISO$-$Oph047 & III & -0.40 & 3.53 & -0.41 & Br$\gamma$ & $<$1.0 & $<$-5.15 & $<$-1.73  & $<$-8.55 \\ 
34 & ISO$-$Oph051 & II & -0.67 & 3.51 & -0.66 & Pa$\beta$ & $<$0.4 & $<$-5.39 & $<$-3.33  & $<$-9.99 \\ 
35 & ISO$-$Oph052 & II & -0.79 & 3.50 & -0.77 & Pa$\beta$ & 5.9 & -4.28 & -1.83  & -8.42 \\ 
36 & ISO$-$Oph053 & II & -0.81 & 3.50 & -0.79 & Pa$\beta$ & $<$0.3 & $<$-5.60 & $<$-3.62  & $<$-10.20 \\ 
37 & ISO$-$Oph055 & III & -1.82 & 3.45 & -1.34 & Pa$\beta$ & $<$1.0 & $<$-5.90 & $<$-4.02  & $<$-10.46 \\ 
38 & ISO$-$Oph056 & II & -0.56 & 3.51 & -0.55 & Pa$\beta$ & $<$0.5 & $<$-5.13 & $<$-2.98  & $<$-9.71 \\ 
39 & ISO$-$Oph058 & III & 0.58 & 3.67 & 0.25 & Pa$\beta$ & $<$0.5 & $<$-4.25 & $<$-1.78  & $<$-9.05 \\ 
40 & ISO$-$Oph059a & II & -0.40 & 3.53 & -0.41 & Br$\gamma$ & $<$0.6 & $<$-5.37 & $<$-1.93  & $<$-8.75 \\ 
41 & ISO$-$Oph059b & II & -0.29 & 3.53 & -0.31 & Pa$\beta$ & $<$3.0 & $<$-4.05 & $<$-1.51  & $<$-8.39 \\ 
42 & ISO$-$Oph062 & II & 0.55 & 3.67 & 0.23 & Pa$\beta$ & 1.4 & -3.79 & -1.15  & -8.41 \\ 
43 & ISO$-$Oph063 & II & -0.73 & 3.50 & -0.73 & Pa$\beta$ & $<$0.3 & $<$-5.53 & $<$-3.52  & $<$-10.13 \\ 
44 & ISO$-$Oph064 & III & 0.04 & 3.57 & -0.04 & Pa$\beta$ & $<$0.7 & $<$-4.51 & $<$-2.14  & $<$-9.20 \\ 
45 & ISO$-$Oph065 & I & -1.35 & 3.47 & -1.10 & Pa$\beta$ & $<$10.0 & $<$-4.56 & $<$-2.20  & $<$-8.69 \\ 
46 & ISO$-$Oph066 & III & -0.67 & 3.51 & -0.66 & Pa$\beta$ & $<$0.8 & $<$-5.01 & $<$-2.81  & $<$-9.47 \\ 
47 & ISO$-$Oph067 & II & 0.21 & 3.59 & 0.04 & Pa$\beta$ & 10.0 & -3.24 & -0.41  & -7.49 \\ 
48 & ISO$-$Oph068 & II & 0.30 & 3.60 & 0.09 & Pa$\beta$ & 0.6 & -4.32 & -1.87  & -8.99 \\ 
49 & ISO$-$Oph069 & III & -0.95 & 3.49 & -0.88 & Pa$\beta$ & $<$0.4 & $<$-5.55 & $<$-3.55  & $<$-10.10 \\ 
50 & ISO$-$Oph070 & II & -3.34 & 3.32 & nan & Br$\gamma$ & $<$1.2 & $<$-6.80 & $<$-3.22  & $<$nan \\ 
51 & ISO$-$Oph072 & II & -0.86 & 3.49 & -0.83 & Pa$\beta$ & 36.0 & -3.55 & -0.83  & -7.39 \\ 
52 & ISO$-$Oph073 & III & 0.57 & 3.67 & 0.24 & Pa$\beta$ & $<$0.4 & $<$-4.32 & $<$-1.88  & $<$-9.15 \\ 
53 & ISO$-$Oph074 & III & -0.25 & 3.54 & -0.28 & Pa$\beta$ & -1.5 & ... & ...  & ... \\ 
54 & ISO$-$Oph075 & II & -0.97 & 3.49 & -0.89 & Br$\gamma$ & $<$3.9 & $<$-4.76 & $<$-1.39  & $<$-7.93 \\ 
55 & ISO$-$Oph076 & II & -0.76 & 3.50 & -0.75 & Br$\gamma$ & $<$3.0 & $<$-4.70 & $<$-1.33  & $<$-7.93 \\ 
56 & ISO$-$Oph078 & II & 0.20 & 3.59 & 0.03 & Pa$\beta$ & $<$0.4 & $<$-4.62 & $<$-2.29  & $<$-9.37 \\ 
57 & ISO$-$Oph079 & II & -1.52 & 3.46 & -1.19 & Pa$\beta$ & $<$3.0 & $<$... & $<$...  & $<$... \\ 
58 & ISO$-$Oph082 & III & -1.99 & 3.44 & -1.43 & Pa$\beta$ & $<$2.0 & $<$-5.71 & $<$-3.77  & $<$-10.19 \\ 
59 & ISO$-$Oph083 & II & -0.04 & 3.56 & -0.12 & Pa$\beta$ & 4.5 & -3.78 & -1.14  & -8.13 \\ 
60 & ISO$-$Oph084a & II & -1.27 & 3.47 & -1.06 & Pa$\beta$ & $<$5.0 & $<$-99.00 & $<$-99.00  & $<$-99.00 \\ 
61 & ISO$-$Oph084b & II & -1.27 & 3.47 & -1.06 & Br$\gamma$ & $<$0.2 & $<$-6.48 & $<$-2.93  & $<$-9.43 \\ 
62 & ISO$-$Oph086 & II & -0.60 & 3.51 & -0.59 & Pa$\beta$ & $<$0.8 & $<$-5.00 & $<$-2.79  & $<$-9.50 \\ 
63 & ISO$-$Oph087 & II & -1.10 & 3.48 & -0.97 & Pa$\beta$ & 2.4 & -4.90 & -2.67  & -9.18 \\ 
64 & ISO$-$Oph088a & II & 0.38 & 3.62 & 0.13 & Pa$\beta$ & 10.5 & -3.05 & -0.15  & -7.30 \\ 
65 & ISO$-$Oph088b & II & 0.41 & 3.62 & 0.14 & Pa$\beta$ & 7.0 & -3.18 & -0.33  & -7.48 \\ 
66 & ISO$-$Oph089a & II & -0.87 & 3.49 & -0.83 & Pa$\beta$ & $<$0.5 & $<$-5.39 & $<$-3.33  & $<$-9.89 \\ 
67 & ISO$-$Oph089b & II & -0.86 & 3.49 & -0.83 & Br$\gamma$ & $<$0.7 & $<$-5.67 & $<$-2.21  & $<$-8.77 \\ 
68 & ISO$-$Oph091 & III & 0.64 & 3.68 & 0.28 & Pa$\beta$ & -0.9 & ... & ...  & ... \\ 
69 & ISO$-$Oph092 & II & 0.91 & 3.70 & 0.35 & Pa$\beta$ & 8.4 & -2.75 & 0.26  & -7.02 \\ 
70 & ISO$-$Oph093a & II & -0.51 & 3.52 & -0.51 & Pa$\beta$ & $<$20.0 & $<$-99.00 & $<$-99.00  & $<$-99.00 \\ 
71 & ISO$-$Oph093b & II & -0.51 & 3.52 & -0.51 & Br$\gamma$ & $<$0.7 & $<$-5.15 & $<$-1.73  & $<$-8.49 \\ 
72 & ISO$-$Oph094 & II & -1.95 & 3.44 & -1.41 & Pa$\beta$ & 5.0 & -5.33 & -3.25  & -9.67 \\ 
73 & ISO$-$Oph095 & II & 0.11 & 3.58 & -0.02 & Pa$\beta$ & $<$0.6 & $<$-4.55 & $<$-2.18  & $<$-9.24 \\ 
74 & ISO$-$Oph096 & III & -0.47 & 3.52 & -0.46 & Pa$\beta$ & -0.9 & ... & ...  & ... \\ 
75 & ISO$-$Oph097 & III & -0.29 & 3.53 & -0.31 & Pa$\beta$ & -1.1 & ... & ...  & ... \\ 
76 & ISO$-$Oph098 & II & -0.06 & 3.56 & -0.14 & Pa$\beta$ & $<$0.7 & $<$-4.62 & $<$-2.28  & $<$-9.26 \\ 
77 & ISO$-$Oph100 & III & -1.76 & 3.45 & -1.32 & Br$\gamma$ & $<$0.7 & $<$-6.21 & $<$-2.69  & $<$-9.14 \\ 
78 & ISO$-$Oph102 & II & -1.43 & 3.47 & -1.14 & Pa$\beta$ & 2.0 & -5.22 & -3.09  & -9.57 \\ 
79 & ISO$-$Oph103 & II & 0.43 & 3.63 & 0.16 & Pa$\beta$ & $<$0.6 & $<$... & $<$...  & $<$... \\ 
80 & ISO$-$Oph104 & III & -1.36 & 3.47 & -1.11 & Br$\gamma$ & $<$0.7 & $<$-5.87 & $<$-2.39  & $<$-8.87 \\ 
81 & ISO$-$Oph105 & II & 0.38 & 3.62 & 0.13 & Pa$\beta$ & 1.7 & -3.81 & -1.18  & -8.33 \\ 
82 & ISO$-$Oph106 & II & -0.49 & 3.52 & -0.48 & Pa$\beta$ & $<$1.0 & $<$-4.82 & $<$-2.55  & $<$-9.33 \\ 
83 & ISO$-$Oph107 & II & -0.58 & 3.51 & -0.57 & Pa$\beta$ & $<$1.3 & $<$... & $<$...  & $<$... \\ 
84 & ISO$-$Oph108 & I & 4.14 & 4.47 & 1.14 & Br$\gamma$ & 1.3 & -1.68 & 1.39  & -6.60 \\ 
85 & ISO$-$Oph109 & III & -0.38 & 3.53 & -0.39 & Br$\gamma$ & $<$0.5 & $<$-5.19 & $<$-1.77  & $<$-8.60 \\ 
86 & ISO$-$Oph110 & II & 0.50 & 3.66 & 0.20 & Pa$\beta$ & $<$0.3 & $<$-4.41 & $<$-2.00  & $<$-9.24 \\ 
87 & ISO$-$Oph111 & III & -0.62 & 3.51 & -0.61 & Pa$\beta$ & $<$1.0 & $<$-4.93 & $<$-2.70  & $<$-9.39 \\ 
88 & ISO$-$Oph112 & II & 0.29 & 3.60 & 0.08 & Pa$\beta$ & 17.9 & ... & ...  & ... \\ 
89 & ISO$-$Oph113a & III & -0.21 & 3.54 & -0.24 & Pa$\beta$ & -4.0 & -99.00 & -99.00  & -99.00 \\ 
90 & ISO$-$Oph113b & III & -0.21 & 3.54 & -0.24 & Br$\gamma$ & -5.6 & -99.00 & -99.00  & -99.00 \\ 
91 & ISO$-$Oph114 & III & 0.92 & 3.71 & 0.36 & Pa$\beta$ & $<$8.0 & $<$... & $<$...  & $<$... \\ 
92 & ISO$-$Oph115 & II & -1.15 & 3.48 & -1.00 & Pa$\beta$ & 2.7 & -4.91 & -2.68  & -9.19 \\ 
93 & ISO$-$Oph116 & II & 0.12 & 3.58 & -0.02 & Pa$\beta$ & $<$1.0 & $<$-4.32 & $<$-1.87  & $<$-8.93 \\ 
94 & ISO$-$Oph117 & II & -0.48 & 3.52 & -0.47 & Pa$\beta$ & 2.4 & -4.44 & -2.03  & -8.82 \\ 
95 & ISO$-$Oph118 & II & -0.38 & 3.53 & -0.39 & Pa$\beta$ & $<$20.0 & $<$... & $<$...  & $<$... \\ 
96 & ISO$-$Oph119 & II & -0.38 & 3.53 & -0.39 & Br$\gamma$ & $<$0.7 & $<$-5.05 & $<$-1.64  & $<$-8.47 \\ 
97 & ISO$-$Oph120 & II & -0.35 & 3.53 & -0.36 & Pa$\beta$ & $<$2.5 & $<$-4.33 & $<$-1.89  & $<$-8.74 \\ 
98 & ISO$-$Oph121a & II & -0.17 & 3.55 & -0.22 & Pa$\beta$ & $<$0.6 & $<$-4.76 & $<$-2.48  & $<$-9.41 \\ 
99 & ISO$-$Oph121b & II & 0.36 & 3.61 & 0.12 & Pa$\beta$ & $<$0.5 & $<$-4.40 & $<$-1.99  & $<$-9.13 \\ 
100 & ISO$-$Oph122 & II & -1.20 & 3.48 & -1.02 & Pa$\beta$ & $<$10.0 & $<$... & $<$...  & $<$... \\ 
101 & ISO$-$Oph123 & II & -1.31 & 3.47 & -1.08 & Pa$\beta$ & 21.9 & -4.10 & -1.57  & -8.06 \\ 
102 & ISO$-$Oph124 & II & -0.24 & 3.54 & -0.27 & Pa$\beta$ & $<$10.0 & $<$... & $<$...  & $<$... \\ 
103 & ISO$-$Oph125 & III & 1.97 & 3.99 & 0.50 & Pa$\beta$ & $<$20.0 & $<$... & $<$...  & $<$... \\ 
104 & ISO$-$Oph126 & III & -0.12 & 3.55 & -0.19 & Pa$\beta$ & $<$0.9 & $<$-4.56 & $<$-2.20  & $<$-9.15 \\ 
105 & ISO$-$Oph127 & II & -0.19 & 3.54 & -0.23 & Br$\gamma$ & $<$0.0 & $<$... & $<$...  & $<$... \\ 
106 & ISO$-$Oph128 & II & 0.27 & 3.60 & 0.08 & Pa$\beta$ & $<$0.4 & $<$-4.58 & $<$-2.23  & $<$-9.34 \\ 
107 & ISO$-$Oph129 & II & 0.06 & 3.58 & -0.04 & Pa$\beta$ & $<$30.0 & $<$... & $<$...  & $<$... \\ 
108 & ISO$-$Oph130 & III & -0.07 & 3.56 & -0.15 & Pa$\beta$ & -0.8 & ... & ...  & ... \\ 
109 & ISO$-$Oph132 & II & 0.57 & 3.67 & 0.24 & Pa$\beta$ & $<$0.9 & $<$-4.01 & $<$-1.46  & $<$-8.73 \\ 
110 & ISO$-$Oph133 & III & 0.23 & 3.59 & 0.06 & Pa$\beta$ & $<$0.9 & $<$-4.28 & $<$-1.83  & $<$-8.93 \\ 
111 & ISO$-$Oph134 & I & 2.17 & 4.11 & 0.56 & Pa$\beta$ & $<$12.0 & $<$... & $<$...  & $<$... \\ 
112 & ISO$-$Oph135 & III & 0.31 & 3.61 & 0.10 & Pa$\beta$ & -1.3 & ... & ...  & ... \\ 
113 & ISO$-$Oph136 & III & -1.28 & 3.47 & -1.06 & Pa$\beta$ & $<$0.9 & $<$... & $<$...  & $<$... \\ 
114 & ISO$-$Oph138 & II & -1.84 & 3.45 & -1.36 & Pa$\beta$ & $<$1.0 & $<$-5.95 & $<$-4.09  & $<$-10.53 \\ 
115 & ISO$-$Oph139a & II & -1.01 & 3.48 & -0.92 & Br$\gamma$ & $<$2.8 & $<$-4.97 & $<$-1.57  & $<$-8.11 \\ 
116 & ISO$-$Oph139b & II & -1.01 & 3.48 & -0.91 & Pa$\beta$ & $<$6.0 & $<$-99.00 & $<$-99.00  & $<$-99.00 \\ 
117 & ISO$-$Oph140 & II & 0.42 & 3.63 & 0.15 & Pa$\beta$ & 3.7 & -3.51 & -0.77  & -7.93 \\ 
118 & ISO$-$Oph141 & I & 2.08 & 4.07 & 0.53 & Br$\gamma$ & 0.9 & -3.11 & 0.10  & -7.52 \\ 
119 & ISO$-$Oph142 & II & -0.04 & 3.56 & -0.12 & Pa$\beta$ & $<$0.5 & $<$-4.72 & $<$-2.42  & $<$-9.41 \\ 
120 & ISO$-$Oph143 & I & 0.69 & 3.68 & 0.30 & Br$\gamma$ & 1.8 & -3.74 & -0.47  & -7.77 \\ 
121 & ISO$-$Oph144 & II & -0.08 & 3.56 & -0.15 & Pa$\beta$ & $<$1.4 & $<$-4.32 & $<$-1.88  & $<$-8.85 \\ 
122 & ISO$-$Oph145 & I & 0.06 & 3.58 & -0.04 & Br$\gamma$ & 2.0 & -4.02 & -0.71  & -7.77 \\ 
123 & ISO$-$Oph146 & III & -0.41 & 3.52 & -0.42 & Br$\gamma$ & $<$0.9 & $<$-4.97 & $<$-1.57  & $<$-8.39 \\ 
124 & ISO$-$Oph147 & II & 0.51 & 3.66 & 0.21 & Pa$\beta$ & $<$0.5 & $<$-4.31 & $<$-1.86  & $<$-9.11 \\ 
125 & ISO$-$Oph148 & III & -0.36 & 3.53 & -0.36 & Pa$\beta$ & $<$0.5 & $<$-5.00 & $<$-2.80  & $<$-9.64 \\ 
126 & ISO$-$Oph149 & III & -0.11 & 3.55 & -0.18 & Pa$\beta$ & -0.9 & ... & ...  & ... \\ 
127 & ISO$-$Oph151 & II & -0.43 & 3.52 & -0.43 & Pa$\beta$ & $<$0.5 & $<$-5.08 & $<$-2.91  & $<$-9.72 \\ 
128 & ISO$-$Oph152 & III & 0.04 & 3.57 & -0.04 & Pa$\beta$ & $<$2.0 & $<$-4.07 & $<$-1.53  & $<$-8.59 \\ 
129 & ISO$-$Oph153 & III & -1.55 & 3.46 & -1.21 & Br$\gamma$ & $<$0.0 & $<$... & $<$...  & $<$... \\ 
130 & ISO$-$Oph154 & II & 0.07 & 3.58 & -0.03 & Pa$\beta$ & $<$1.0 & $<$-4.38 & $<$-1.95  & $<$-9.01 \\ 
131 & ISO$-$Oph155 & II & 0.57 & 3.67 & 0.24 & Pa$\beta$ & 4.0 & -3.30 & -0.49  & -7.76 \\ 
132 & ISO$-$Oph156 & III & -0.77 & 3.50 & -0.77 & Pa$\beta$ & $<$0.4 & $<$-5.43 & $<$-3.38  & $<$-9.98 \\ 
133 & ISO$-$Oph157 & III & -1.10 & 3.48 & -0.97 & Br$\gamma$ & $<$1.0 & $<$-5.76 & $<$-2.29  & $<$-8.80 \\ 
134 & ISO$-$Oph158 & III & -1.59 & 3.46 & -1.23 & Pa$\beta$ & $<$1.0 & $<$-5.74 & $<$-3.81  & $<$-10.27 \\ 
135 & ISO$-$Oph159 & I & 0.16 & 3.58 & 0.01 & Br$\gamma$ & $<$0.0 & $<$... & $<$...  & $<$... \\ 
136 & ISO$-$Oph160 & II & -1.73 & 3.45 & -1.30 & Pa$\beta$ & 3.3 & -5.26 & -3.15  & -9.60 \\ 
137 & ISO$-$Oph161 & II & -0.01 & 3.56 & -0.11 & Br$\gamma$ & $<$0.8 & $<$-4.67 & $<$-1.30  & $<$-8.30 \\ 
138 & ISO$-$Oph163 & II & 0.26 & 3.60 & 0.07 & Pa$\beta$ & 2.0 & -3.87 & -1.27  & -8.37 \\ 
139 & ISO$-$Oph164 & II & -1.47 & 3.46 & -1.16 & Pa$\beta$ & 0.8 & -5.63 & -3.66  & -10.13 \\ 
140 & ISO$-$Oph165 & II & -1.34 & 3.47 & -1.10 & Pa$\beta$ & 9.6 & -4.56 & -2.20  & -8.69 \\ 
141 & ISO$-$Oph166 & II & 0.22 & 3.59 & 0.05 & Pa$\beta$ & 3.3 & -3.64 & -0.95  & -8.05 \\ 
142 & ISO$-$Oph167 & I & 2.27 & 4.14 & 0.58 & Pa$\beta$ & $<$4.0 & $<$-2.01 & $<$1.27  & $<$-6.43 \\ 
143 & ISO$-$Oph168 & II & 0.11 & 3.58 & -0.02 & Pa$\beta$ & 1.3 & -4.19 & -1.69  & -8.75 \\ 
144 & ISO$-$Oph169 & III & -0.26 & 3.54 & -0.29 & Pa$\beta$ & -0.7 & ... & ...  & ... \\ 
144 & ISO$-$Oph169 & III & -0.77 & 3.50 & -0.76 & Pa$\beta$ & $<$0.9 & $<$-5.06 & $<$-2.88  & $<$-9.48 \\ 
145 & ISO$-$Oph169 & III & -0.26 & 3.54 & -0.29 & Pa$\beta$ & -0.7 & ... & ...  & ... \\ 
145 & ISO$-$Oph169 & III & -0.77 & 3.50 & -0.76 & Pa$\beta$ & $<$0.9 & $<$-5.06 & $<$-2.88  & $<$-9.48 \\ 
146 & ISO$-$Oph170 & II & -3.34 & 3.32 & nan & Pa$\beta$ & $<$3.0 & $<$-6.68 & $<$-5.09  & $<$nan \\ 
147 & ISO$-$Oph171 & II & -0.78 & 3.50 & -0.77 & Pa$\beta$ & $<$4.0 & $<$... & $<$...  & $<$... \\ 
148 & ISO$-$Oph172 & II & -0.60 & 3.51 & -0.59 & Pa$\beta$ & $<$1.0 & $<$-4.91 & $<$-2.68  & $<$-9.38 \\ 
149 & ISO$-$Oph173 & III & -0.21 & 3.54 & -0.25 & Pa$\beta$ & $<$3.0 & $<$... & $<$...  & $<$... \\ 
150 & ISO$-$Oph174 & III & -0.96 & 3.49 & -0.89 & Pa$\beta$ & -0.7 & ... & ...  & ... \\ 
151 & ISO$-$Oph175 & II & -1.86 & 3.45 & -1.36 & Pa$\beta$ & 15.7 & -4.76 & -2.47  & -8.91 \\ 
152 & ISO$-$Oph176a & II & -1.22 & 3.47 & -1.03 & Pa$\beta$ & $<$0.5 & $<$-5.65 & $<$-3.68  & $<$-10.18 \\ 
153 & ISO$-$Oph176b & II & -1.21 & 3.48 & -1.02 & Br$\gamma$ & $<$0.6 & $<$-6.02 & $<$-2.52  & $<$-9.02 \\ 
154 & ISO$-$Oph177 & II & -0.43 & 3.52 & -0.43 & Pa$\beta$ & $<$0.4 & $<$-5.13 & $<$-2.97  & $<$-9.78 \\ 
155 & ISO$-$Oph178 & II & -0.73 & 3.50 & -0.73 & Pa$\beta$ & 2.0 & -4.72 & -2.41  & -9.03 \\ 
156 & ISO$-$Oph179 & III & -0.88 & 3.49 & -0.84 & Br$\gamma$ & $<$0.7 & $<$-5.68 & $<$-2.21  & $<$-8.77 \\ 
157 & ISO$-$Oph180 & III & 0.31 & 3.61 & 0.10 & Pa$\beta$ & -6.3 & ... & ...  & ... \\ 
158 & ISO$-$Oph181 & III & -1.30 & 3.47 & -1.07 & Pa$\beta$ & -5.5 & ... & ...  & ... \\ 
159 & ISO$-$Oph182 & I & 0.22 & 3.59 & 0.05 & Br$\gamma$ & 5.4 & -3.41 & -0.17  & -7.26 \\ 
160 & ISO$-$Oph183 & III & -0.18 & 3.54 & -0.23 & Pa$\beta$ & $<$0.7 & $<$-4.72 & $<$-2.42  & $<$-9.34 \\ 
161 & ISO$-$Oph184 & III & 0.22 & 3.59 & 0.05 & Pa$\beta$ & -0.7 & ... & ...  & ... \\ 
162 & ISO$-$Oph185 & II & -1.08 & 3.48 & -0.96 & Pa$\beta$ & $<$0.2 & $<$-5.96 & $<$-4.11  & $<$-10.62 \\ 
163 & ISO$-$Oph186 & III & -1.13 & 3.48 & -0.98 & Pa$\beta$ & $<$1.5 & $<$-5.16 & $<$-3.02  & $<$-9.53 \\ 
164 & ISO$-$Oph187 & II & -0.56 & 3.51 & -0.55 & Pa$\beta$ & 5.6 & -4.12 & -1.61  & -8.34 \\ 
165 & ISO$-$Oph188 & III & -0.17 & 3.54 & -0.22 & Pa$\beta$ & -0.5 & ... & ...  & ... \\ 
166 & ISO$-$Oph189 & III & -0.86 & 3.49 & -0.83 & Pa$\beta$ & $<$1.5 & $<$-4.94 & $<$-2.72  & $<$-9.28 \\ 
167 & ISO$-$Oph190 & II & -1.65 & 3.46 & -1.26 & Pa$\beta$ & $<$5.0 & $<$... & $<$...  & $<$... \\ 
168 & ISO$-$Oph191 & III & -0.03 & 3.56 & -0.12 & Pa$\beta$ & $<$0.5 & $<$-4.74 & $<$-2.45  & $<$-9.44 \\ 
169 & ISO$-$Oph192 & III & -1.06 & 3.48 & -0.94 & Pa$\beta$ & $<$2.0 & $<$-4.98 & $<$-2.77  & $<$-9.29 \\ 
170 & ISO$-$Oph193 & II & -1.22 & 3.47 & -1.03 & Pa$\beta$ & 1.8 & -5.20 & -3.08  & -9.58 \\ 
171 & ISO$-$Oph194 & II & -0.73 & 3.50 & -0.73 & Pa$\beta$ & $<$0.5 & $<$-5.30 & $<$-3.21  & $<$-9.83 \\ 
172 & ISO$-$Oph195 & II & -0.10 & 3.55 & -0.17 & Pa$\beta$ & 0.8 & -4.53 & -2.17  & -9.13 \\ 
173 & ISO$-$Oph196 & II & -0.67 & 3.51 & -0.66 & Pa$\beta$ & 3.2 & -4.44 & -2.03  & -8.70 \\ 
174 & ISO$-$Oph197 & III & -1.11 & 3.48 & -0.97 & Pa$\beta$ & $<$1.0 & $<$-5.29 & $<$-3.19  & $<$-9.71 \\ 
175 & ISO$-$Oph198 & III & 0.73 & 3.69 & 0.31 & Pa$\beta$ & $<$0.4 & $<$-4.12 & $<$-1.60  & $<$-8.90 \\ 
176 & ISO$-$Oph199 & II & -0.13 & 3.55 & -0.19 & Pa$\beta$ & 1.7 & -4.23 & -1.75  & -8.70 \\ 
\hline 
\end{longtable} 
\noindent 
$M_\star$ and age are derived using the evolutionary tracks of \citet{Baraffe98} and assuming a single isochrone at 1 Myr as explained in \citet{Rigliaco11a}.

\end{document}